\documentclass[twocolumn]{aastex62}

\usepackage{multirow}

\def\JB{{\rm Jy~beam^{-1}}}
\def\mJB{{\rm mJy~beam^{-1}}}

\def\mJ{{\rm mJy}}
\def\kms{{\rm km~s^{-1}}}
\def\Ms{M_{\sun}}
\def\Ls{L_{\sun}}

\def\II{{\rm I\hspace{-0.1em}I}}

\graphicspath{{./}{figures/}}

\accepted{October 24, 2019}
\submitjournal{ApJ}

\shorttitle{Protostellar evolution in Serpens Main}
\shortauthors{Aso et al.}

\begin{document}

\title{Protostellar Evolution in Serpens Main: Possible Origin of Disk-Size Diversity}

\correspondingauthor{Yusuke Aso}
\email{yaso@asiaa.sinica.edu.tw}

\author[0000-0002-8238-7709]{Yusuke Aso}
\affil{Academia Sinica Institute of Astronomy and Astrophysics, 11F of ASMA Building, No.1, Sec. 4, Roosevelt Rd, Taipei 10617, Taiwan}

\author[0000-0001-9304-7884]{Naomi Hirano}
\affil{Academia Sinica Institute of Astronomy and Astrophysics, 11F of ASMA Building, No.1, Sec. 4, Roosevelt Rd, Taipei 10617, Taiwan}

\author[0000-0003-3283-6884]{Yuri Aikawa}
\affil{Department of Astronomy, Graduate School of Science, The University of Tokyo, 7-3-1 Hongo, Bunkyo-ku, Tokyo 113-0033, Japan}

\author[0000-0002-0963-0872]{Masahiro N. Machida}
\affil{Department of Earth and Planetary Sciences, Faculty of Sciences Kyushu University, Fukuoka 812-8581, Japan}

\author[0000-0003-0998-5064]{Nagayoshi Ohashi}
\affil{Academia Sinica Institute of Astronomy and Astrophysics, 11F of ASMA Building, No.1, Sec. 4, Roosevelt Rd, Taipei 10617, Taiwan}
\affil{Subaru Telescope, National Astronomical Observatory of Japan 650 North A'ohoku Place, Hilo, HI 96720, USA}

\author[0000-0003-0769-8627]{Masao Saito}
\affil{Nobeyama Radio Observatory, Nobeyama, Minamimaki, Minamisaku, Nagano 384-1305, Japan}
\affil{SOKENDAI, Department of Astronomical Science, Graduate University for Advanced Studies}

\author[0000-0003-0845-128X]{Shigehisa Takakuwa}
\affil{Academia Sinica Institute of Astronomy and Astrophysics, 11F of ASMA Building, No.1, Sec. 4, Roosevelt Rd, Taipei 10617, Taiwan}
\affil{Department of Physics and Astronomy, Graduate School of Science and Engineering, Kagoshima University, 1-21-35 Korimoto, Kagoshima, Kagoshima 890-0065, Japan}

\author[0000-0003-1412-893X]{Hsi-Wei Yen}
\affil{Academia Sinica Institute of Astronomy and Astrophysics, 11F of ASMA Building, No.1, Sec. 4, Roosevelt Rd, Taipei 10617, Taiwan}

\author[0000-0001-5058-695X]{Jonathan P. Williams}
\affil{Institute for Astronomy, University of Hawaii at Manoa, Honolulu, Hawaii, USA}

\begin{abstract}
We have observed the submillimeter continuum condensations SMM2, SMM4, SMM9, and SMM11 in the star forming cluster Serpens Main using the Atacama Large Millimeter/submillimeter Array during Cycle 3 in the 1.3 mm continuum, $^{12}$CO $J=2-1$, SO $J_N=6_5-5_4$, and C$^{18}$O $J=2-1$ lines at an angular resolution of $\sim 0\farcs 55$ (240 au). Sixteen sources have been detected in the 1.3 mm continuum, which can be classified into three groups. Group 1 consists of six sources showing extended continuum emission and bipolar/monopolar $^{12}$CO outflows. Although all the Group 1 members are classified as Class 0 protostars, our observations suggest evolutionary trends among them in terms of $^{12}$CO outflow dynamical time, SO emission distribution, C$^{18}$O fractional abundance, and continuum morphology. Group 2 consists of four sources associated with a continuum filamentary structure and no $^{12}$CO outflows. Central densities estimated from the 1.3 mm continuum intensity suggest that they are prestellar sources in a marginally Jeans unstable state. Group 3 consists of six Spitzer sources showing point-like 1.3 mm continuum emission and clumpy $^{12}$CO outflows. These features of Group 3 suggest envelope dissipation, preventing disk growth from the present size, $r\lesssim 60$ au. The Group 3 members are protostars that may be precursors to the T Tauri stars associated with small disks at tens-au radii identified in recent surveys. 
\end{abstract}



\section{Introduction} \label{sec:intro}
Protoplanetary disks were identified around T Tauri stars or Class $\II$ sources in early studies \citep[reviewed by][]{wi.ci11}, and Keplerian disks have also been identified around Class I \citep{taka12, hars14, aso15} and Class 0 \citep[e.g.,][]{muri13, lee14, aso17b} protostars. A typical picture of such circumstellar disks, a radius of $\sim 100$ au, was built up by the millimeter images of the disk around the Class I protostar HL Tau in the Taurus star forming region observed with ALMA \citep{alma15}. A disk survey using ALMA, DSHARP, at 1.25 mm also supports this picture with a sample of 20 Class $\II$ disks \citep{andr18}.

The formation process of disks have been discussed in recent studies. A classical scenario suggested by \citet{tere84} argued that a core in rigid-body rotation supplies angular momentum inward, and a disk slowly forms receiving little angular momentum at early stages, then grows rapidly receiving the majority of its angular momentum at late stages. Another scenario suggested by \citet{basu98} starts from a rotating magnetized core with a uniform rotational velocity. This initial distribution of the rotational velocity produces more rapid disk formation at an early stage and slower growth at a later stage than the classical scenario by \citet{tere84}. With a sample of 18 protostellar disks, \citet{yen17} investigated the relation between disk radius versus central stellar mass and age. The relation suggests that 100-au sized disks form at the youngest protostellar (Class 0) stage rapidly, then growth of the disks slows down at the Class I stage. This result supports the rapid-formation slow-growth scenario. Theoretical studies have also attempted to form 100-au sized disks against magnetic braking by considering, for example, appropriate numerical setting \citep{mach14} or non-ideal magnetohydrodynamics (MHD) effects.

The observations described above have shown circumstellar disks with $r\sim 100$ au, as well as formation and growth scenarios of such disks. However, recent disk ALMA surveys of Lupus and Ophiuchus show that most Spitzer-selected protoplanetary disks have radii smaller than 30~au \citep{an.wi16, ciez19}.
Another recent observational study with 33 Class $\II$ disks also reported that the radii of these disks range from a few tens to several hundreds au in continuum and molecular line emission at (sub)mm wavelengths regardless of their age \citep{na.be18}. Because these disks are likely in their final stages of growth, these observational results imply that not all disks grow up to 100 au in radius, and such a 100-au disk is no longer considered typical. Such diversity in disk sizes at the T Tauri stage might originate in evolutionary process at the earlier, i.e., protostellar stage.


In order to study the relation between protostellar evolution and the size diversity of protoplanetary disks in the T Tauri phase, we observed the star forming cluster Serpens Main with the Atacama Large Millimeter/submillimeter Array (ALMA) in the $^{12}$CO $J=2-1$ (230.538 GHz), SO $J_N=6_5-5_4$ (219.949 GHz), and C$^{18}$O $J=2-1$ (219.560 GHz) lines and the 1.3 mm continuum. Serpens Main, extending over a scale of $10^4$ au, is a good target for this purpose because it has a high number density of protostars and a high protostellar fraction \citep{li19}, suggesting that this region is in an early evolutionary state with active disk formation. The distance to Serpens Main is 429 pc \citep{dzib11}. Our targets are the submillimeter condensations SMM2, SMM4, SMM9, and SMM11 identified with the James Clerk Maxwell Telescope (JCMT) \citep{davi99}.

\begin{figure}[ht!]
\epsscale{1.2}
\plotone{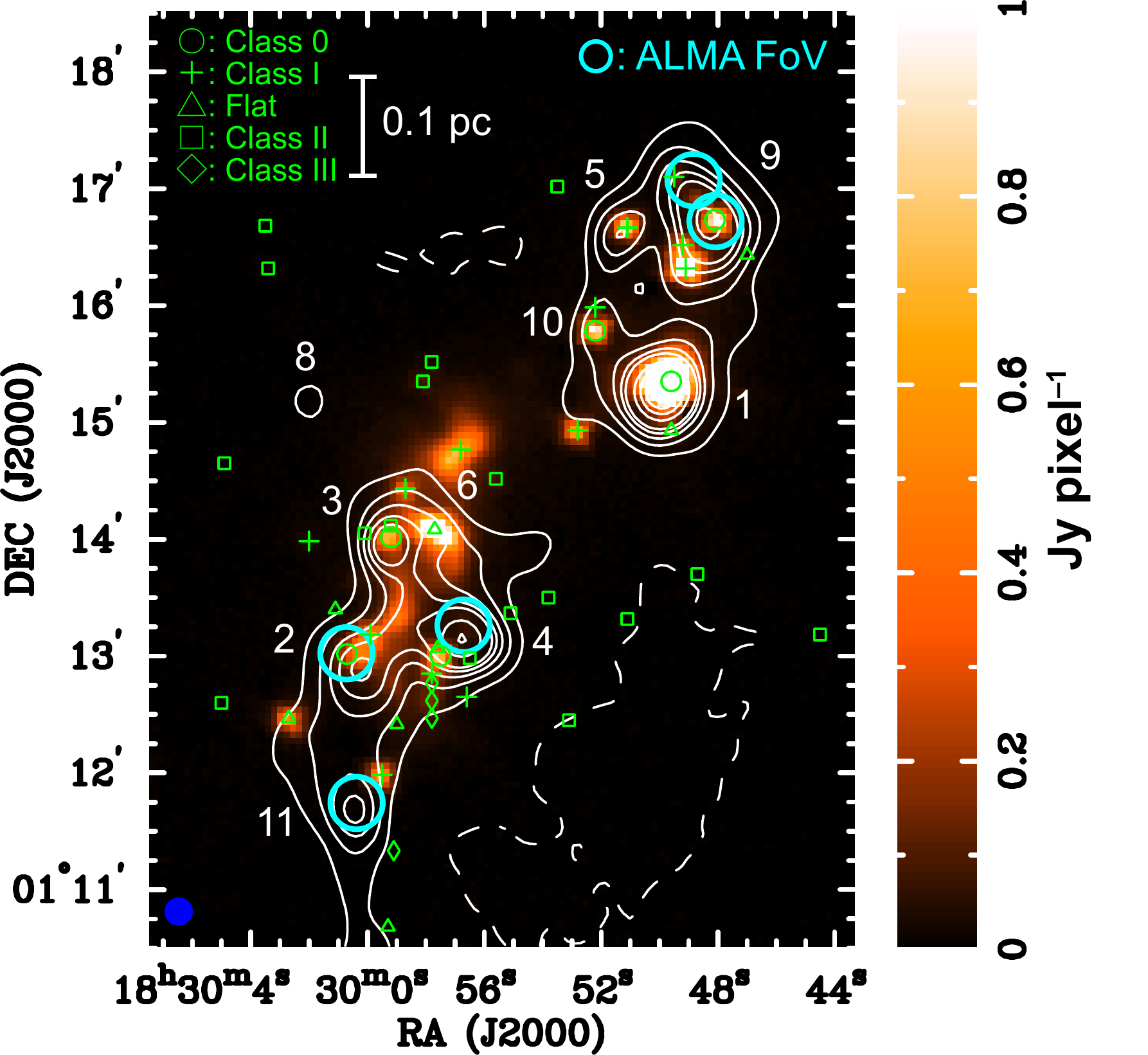}
\caption{Serpens Main at JCMT 850 $\micron$ (contours) and Herschel 70 $\micron$ (color). Contour levels are from $3\sigma$ to $15\sigma$ in $3\sigma$ steps and from $15\sigma$ in $15\sigma$ steps, where $1\sigma$ corresponds to $0.15\ \JB$. The pixel size of the Herschel image is $3\farcs 2$. The green marks indicate YSO positions identified by Spitzer observations \citep{dunh15}. Submillimeter continuum condensations \citep{davi99} are labeled with numbers. The cyan circles indicate FWHM of the primary beams in our ALMA observations. The filled ellipse at the bottom left corner denotes the JCMT beam size ($14\arcsec$).
\label{fig:large}}
\end{figure}

Figure \ref{fig:large} shows an entire map of Serpens Main including the submillimeter condensations. Serpens Main consists of the NW and SE subclusters, as shown in the 850 $\micron$ map. The Class 0 protostar S68N is the 70 $\micron$ source at the center of the westernmost ALMA field of view (FoV) near ``9" in Figure \ref{fig:large}. SMM9 is elongated from S68N to the northeast at 850 $\micron$ (Figure \ref{fig:large}). SMM11 is located in a 0.1 pc scale filament extended from the SE subcluster to the south identified by observations in N$_2$H$^+$ $J=1-0$ using the Combined Array for Research in Millimeter-wave Astronomy (CARMA) \citep{leek14}. Spitzer observations identified Class 0 protostars in SMM2, SMM4, and SMM9, and Class I protostars in SMM2, SMM9, and SMM11 \citep{dunh15}. In addition, one Class 0 protostar was identified in SMM11 by recent observations using CARMA \citep{leek14} and ALMA \citep{aso17a} at millimeter wavelengths. The ALMA observations also identified two other Class 0 protostars, SMM4A and SMM4B, revealing a $\sim$500 au sized disk around SMM4A \citep{aso18}. JCMT observations found several $^{12}$CO outflows over the entire region of Serpens Main \citep{davi99}. The fractional abundance of C$^{18}$O relative to H$_2$ is also estimated from JCMT and Institut de radioastronomie millim\'etrique (IRAM) 30 m observations to be 2-10$\times 10^{-8}$ in this region, which is $\sim 10$ times lower than the canonical interstellar value \citep{du-ca10}.

The outline of this paper is the following. Our ALMA observations and data reduction are described in Section \ref{sec:obs}. In Section \ref{sec:res}, we present the results of the continuum and molecular lines derived from the ALMA observations in sixteen sources, and divide them into Group 1, 2, and 3. The spectral energy distributions (SEDs) of two Group 1 sources are also shown here. Further analyses are performed in Section \ref{sec:ana} to investigate morphology and velocity structures of the Group 1 outflows, Jeans instability of the Group 2 members, and the fractional abundance of C$^{18}$O for all the groups. Evolutionary trends in the Group 1 members and nature of Group 3 will be discussed in Section \ref{sec:dis}. We present a summary of the results in Section \ref{sec:conc}.

\section{ALMA Observations} \label{sec:obs}
We observed five regions, highlighted by the cyan circles in Figure \ref{fig:large}, in Serpens Main using ALMA in Cycle 3 on 2016 May 19 and 21. The results of the SMM11 and SMM4 condensations were reported in \citet{aso17a} and \citet{aso18}, respectively. This paper reports comprehensive results including all the detected sources. 
On-source observing time for each field of view except SMM4 was 4.5 and 9.0 min in the first and the second days, respectively; that for SMM4 was 4.5 and 10.5 min, respectively. The numbers of 12 m antennas were 37 and 39 in the first and the second days, respectively, and the antenna configuration of the second day was more extended than that of the first day. The minimum projected baseline length was 15 m. This minimum projected baseline limits the response of our observations; if observed emission is a Gaussian component with a FWHM of $8\farcs0$ (3400 au), $\sim 50$\% of its flux is missed. \citep{wi.we94}. Spectral windows for the $^{12}$CO $(J=2-1)$, C$^{18}$O ($J=2-1$), and SO ($J_N=6_5 -5_4$) lines have 3840, 1920, and 960 channels covering 117, 59 and 59 MHz band widths at frequency resolutions of 30.5, 30.5, and 61.0 kHz, respectively. When maps are generated, 32, 2, and 4 channels are binned for the $^{12}$CO, C$^{18}$O, and SO lines and the resultant velocity resolutions are 1.27, 0.083, and 0.33 $\kms$, respectively. Two other spectral windows covering 216-218 GHz and 232-234 GHz were assigned to the continuum emission.

All imaging was carried out with Common Astronomical Software Applications (CASA); the CASA version for the calibration procedure is 4.7.0. The visibilities were Fourier transformed and CLEANed with a Briggs robust parameter of 0.0 and a threshold of 3$\sigma$ for all the lines and continuum data. Multi-scale CLEAN was adopted, where CLEAN components were point sources or Gaussian sources with a FWHM of $\sim 1\farcs5$.

We also performed self-calibration for the continuum data of the fields of SMM11 and SMM4 using tasks in CASA ({\it clean}, {\it gaincal}, and {\it applycal}).
Only the phase was calibrated first with the solution interval of 3 scans ($\sim 18$s). Then, using the derived table, the amplitude and the phase were calibrated together. Successful self-calibration improved the rms noise level of the continuum maps by a factor of $\sim 2$. The obtained calibration tables for the continuum data were also applied to the line data. We did not adopt the self-calibrated results for the other three fields because the rms noise level was not improved significantly in those fields. The noise level of the line maps were measured in emission-free channels. The precision of absolute flux is $\sim 10$\% at Band 6 of ALMA. The parameters of our observations are summarized in Table \ref{tab:obs}.

\begin{deluxetable*}{lllll}
\tablecaption{Summary of the ALMA observational parameters \label{tab:obs}}
\tablehead{
\colhead{Date} & \multicolumn{4}{l}{2016.May.19, 21 (project ID: 2015.1.01478.S)}\\
\colhead{Projected baseline length} & \multicolumn{4}{l}{15 - 613 m (11 - 460 k$\lambda$)}\\
\colhead{Primary beam} & \multicolumn{4}{l}{$27\arcsec$}\\
\colhead{Bandpass calibrator} & \multicolumn{4}{l}{J1751$+$0939}\\
\colhead{Amplitude calibrator} & \multicolumn{4}{l}{Titan}\\
\colhead{Phase calibrator} & \multicolumn{4}{l}{J1830$+$0619 (470 mJy), J1824$+$0119 (79 mJy)}\\
\colhead{Coordinate centers (J2000)} & \multicolumn{4}{l}{$18^{\rm h}29^{\rm m}48\fs 08$, $01\arcdeg 16\arcmin 43\farcs 30$ (S68N)}\\
\colhead{} & \multicolumn{4}{l}{$18^{\rm h}29^{\rm m}48\fs 83$, $01\arcdeg 17\arcmin 04\farcs 30$ (S68Nbc)}\\
\colhead{} & \multicolumn{4}{l}{$18^{\rm h}30^{\rm m}00\fs 72$, $01\arcdeg 13\arcmin 01\farcs 40$ (SMM2)}\\
\colhead{} & \multicolumn{4}{l}{$18^{\rm h}29^{\rm m}56\fs 71$, $01\arcdeg 13\arcmin 15\farcs 60$ (SMM4)}\\
\colhead{} & \multicolumn{4}{l}{$18^{\rm h}30^{\rm m}00\fs 38$, $01\arcdeg 11\arcmin 44\farcs 55$ (SMM11)}
}
\startdata
 & Continuum & $^{12}$CO ($J=2-1$) & SO ($J_N = 6_5 - 5_4$) & C$^{18}$O ($J=2-1)$ \\
\hline
Frequency (GHz) & 225 & 230.538000 & 219.949433 & 219.560358 \\
Bandwidth/velocity resolution & 4 GHz & $1.27\ \kms$ & $0.33\ \kms$ & $0.083\ \kms$ \\
Beam (P.A.) & $0\farcs 57\times 0\farcs46\ (-86\arcdeg)$ & $0\farcs 61\times 0\farcs 51\ (-82\arcdeg)$ & $0\farcs65\times 0\farcs 52\ (-85\arcdeg)$ & $0\farcs 65\times 0\farcs 52\ (-85\arcdeg)$ \\
rms noise level ($\mJB$) & 0.1 & 3 & 7 & 10 \\
\enddata
\end{deluxetable*}

\section{Results} \label{sec:res}
Sixteen sources were detected in the 1.3 mm continuum in our ALMA observations. The sources in the SMM2, SMM4, and SMM11 condensations are labeled using the names of the associated condensations, while those in SMM9 are labeled according to the names of 3 mm sources such as S68N, S68Nb, and S68Nc, identified by the Berkeley-Illinois-Maryland Association (BIMA) interferometer \citep{wi.my00}. All the condensations identified with the JCMT and the BIMA interferometer, except S68N, have been spatially resolved into multiple components by our ALMA observations. 

The spatial distributions of the 1.3 mm and $^{12}$CO emissions vary from source to source. On the basis of these properties, we divide the sixteen sources into three groups in this paper as shown in Table \ref{tab:coord}. The six sources belonging to Group 1 appear to be typical protostars including the three Class 0 protostars reported in our previous works \citep{aso18, aso17a}. These six sources have extended components ($>1000$ au) in 1.3 mm continuum emission and $^{12}$CO bipolar/monopolar outflows. The four sources categorized as Group 2, are located in a filamentary structure of 1.3 mm continuum emission, and have no outflow in the $^{12}$CO emission. The other six sources belonging to Group 3 show point-like 1.3 mm continuum emission and compact $^{12}$CO emission. The results of each group are discussed in more detail in the following subsections.

\begin{deluxetable*}{cccccccc}
\tablecaption{Names, coordinates, and Classes of the 1.3 mm sources detected with our ALMA observations. The bolometric luminosity $L_{\rm bol}$, bolometric temperature $T_{\rm bol}$, and evolutionary class are cited from \citet{dunh15, aso17a, aso18}, or derived in Section \ref{sec:sed}. 
\label{tab:coord}}
\tablehead{
\colhead{ } & SSTc2d ID & \colhead{$\alpha$ (J2000)} & \colhead{$\delta$ (J2000)} & $L_{\rm bol}$ & $T_{\rm bol}$ & Class & Group in\\
\colhead{ } & \colhead{ } & \colhead{h:m:s} & \colhead{d:m:s} & \colhead{($\Ls$)} & \colhead{(K)} & \colhead{ } & \colhead{this paper}
}
\startdata
SMM4A &  & 18:29:56.72 & 01:13:15.6 & \multirow{2}{*}{$<$2.6\tablenotemark{a}} & \multirow{2}{*}{$<$30\tablenotemark{a}} & 0 & 1\\
SMM4B &  & 18:29:56.53 & 01:13:11.5 &  &  & 0 & 1\\
S68N & J182948.1+011644 & 18:29:48.09 & 01:16:43.3 & 14 & 30 & 0 & 1\\
S68Nc1 &  & 18:29:48.72 & 01:16:55.6 & $<$2.1 & $<$40 & 0 & 1\\
S68Nb1 &  & 18:29:49.51 & 01:17:10.9 & $<$0.9 & $<$60 & 0 & 1\\
SMM11 &  & 18:30:00.39 & 01:11:44.6 & $<$0.9 & $<$29 & 0 & 1\\
\hline
S68Nc2 &  & 18:29:48.98 & 01:17:07.3 & - & - & - & 2\\
S68Nc3 &  & 18:29:48.85 & 01:17:04.4 & - & - & - & 2\\
S68Nc4 &  & 18:29:48.88 & 01:17:03.1 & - & - & - & 2\\
S68Nc5 &  & 18:29:48.68 & 01:17:02.3 & - & - & - & 2\\
\hline
S68Nb2 & J182949.5+011706 & 18:29:49.60 & 01:17:05.7 & 1.2 & 570 & I & 3\\
SMM2A & J183000.7+011301 & 18:30:00.74 & 01:12:56.2 & \multirow{2}{*}{8\tablenotemark{a}} & \multirow{2}{*}{29\tablenotemark{a}} & 0 & 3\\
SMM2B & J183000.7+011301 & 18:30:00.67 & 01:13:00.1 &  &  & 0 & 3\\
SMM2C & J182959.9+011311 & 18:29:59.94 & 01:13:11.3 & 7 & 120 & I & 3\\
SMM11B & J182959.5+011159 & 18:29:59.62 & 01:11:59.5 & \multirow{2}{*}{15\tablenotemark{a}} & \multirow{2}{*}{120\tablenotemark{a}} & I & 3\\
SMM11C & J182959.5+011159 & 18:29:59.59 & 01:11:58.2 &  &  & I & 3\\
\enddata
\tablenotetext{a}{Because the pairs of SMM4A and 4B, SMM2A and 2B, and SMM11B and 11C were not spatially resolved at far-infrared or submillimeter wavelengths, their bolometric luminosities and bolometric temperatures are the value for each pair.}
\end{deluxetable*}

\subsection{Group 1} \label{sec:cont}
\begin{figure*}[ht!]
\epsscale{1.1}
\plotone{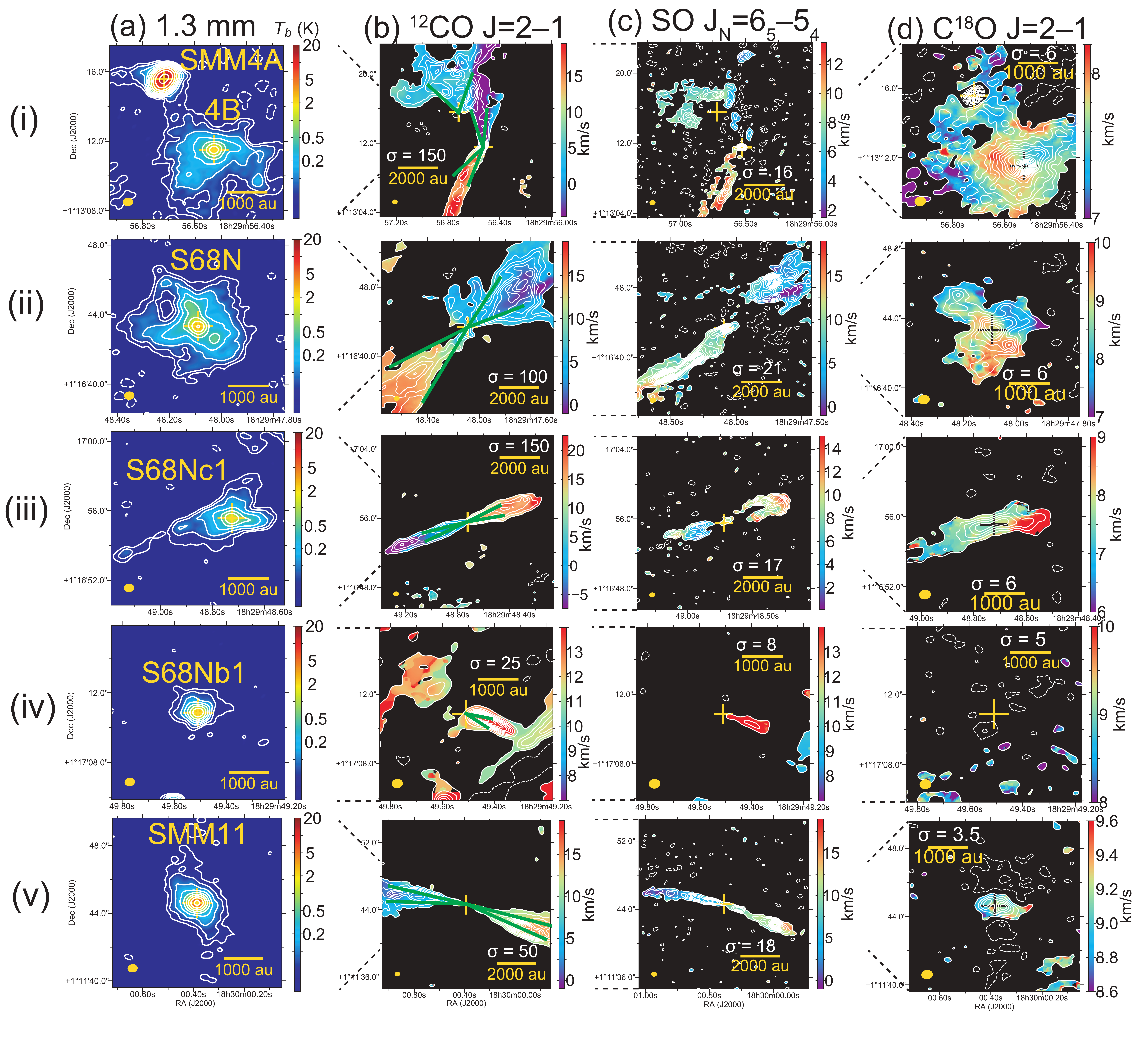}
\caption{Continuum (a) and line (b-d) maps of Group 1.
In the line maps, the contour and color maps show integrated intensity (moment 0) and mean velocity (moment 1) maps, respectively. The $1\sigma$ level in the unit of $\mJB~\kms$ are written in each panel. (a) The 1.3 mm continuum. Contour levels are $3,6,12,24,\dots \times \sigma$, where 1$\sigma$ corresponds to $0.1\ \mJB$.
(b) The $^{12}$CO $J=2-1$ line. Contour levels are in $5\sigma$ steps. The V-shaped green lines denote the intensity-weighted outflow opening angles measured in Section \ref{sec:flow}. The integrated velocity ranges are (i) $-40$ - 56, (ii) $-15$ - 30, (iii) $-45$ - 45, (iv) 9 - 18 , and (v) $-8$ - 23 $\kms$.
(c) The SO $J_N=6_5-5_4$ line. Contour levels are in $3\sigma$ steps. The integrated velocity ranges are (i) 2 - 14, (ii) $-7$ - 21, (iii) $-1$ - 17, (iv) 8 - 10 plus 14 - 16, and (v) $-2$ - 18 $\kms$.
(d) The C$^{18}$O $J=2-1$ line. Contour levels are in $3\sigma$ steps. The integrated velocity ranges are (i) 6.0 - 9.0 (ii) 6.0 - 8.0 plus 8.8 - 10.4, (iii) 6.2 - 8.2 plus 8.9 - 10.1, (iv) 6.3 - 10.1, and (v) 8.3 - 9.9 $\kms$.
The plus signs denote the peak positions of the 1.3 mm continuum emission as derived from 2D Gaussian fittings. The filled ellipses at the bottom-left corners denote the ALMA synthesized beams. 
\label{fig:gr1}}
\end{figure*}

Figure \ref{fig:gr1}a shows 1.3 mm continuum images of the Group 1 members. All of the Group 1 members show components as large as $\sim 1000$ au or more extended in the 1.3 mm continuum.
Peak brightness temperatures $T_b$ of the 1.3 mm continuum emission range from $\sim2$ K to $\sim20$ K. The highest $T_b$, $\sim 20$ K, is measured in SMM4A. This value is as high as a gas temperature estimated from CARMA observations at a spatial resolution of 3000 au \citep{leek14}. This indicates that the 1.3 mm emission is optically thick in SMM4A.
Peak positions were determined by 2D Gaussian fittings to the continuum images. The deconvolved FWHM listed in Table \ref{tab:info} were also derived from the Gaussian fittings. The peak intensity $I_{\rm 1.3mm}^{\rm peak}$ and total flux density $F_{\rm 1.3mm}$ were measured within the $3\sigma$ contours enclosing each source after primary beam correction. 
For the spatially resolved emission, the uncertainty of each flux density was calculated from the rms noise level of intensity $\sigma _i$ in the unit of Jy beam$^{-1}$ and the integrated area $\Omega$ in the unit of beam as $\sigma_i\sqrt{4\Omega}$, where the factor 4 is due to the Nyquist sampling.
The measured parameters are listed in Table \ref{tab:info}. S68N was identified as a Class 0 protostar by Spitzer observations \citep{dunh15}. SMM11, SMM4A, and SMM4B were also reported as Class 0 protostars in previous observational studies using ALMA \citep{aso17a, aso18}. S68Nb1 and S68Nc1 are also identified as Class 0 protostars based on their SEDs, as inspected in Section \ref{sec:sed}.

Figure \ref{fig:gr1}b shows integrated intensity (moment 0) and mean velocity (moment 1) maps of the $^{12}$CO $J=2-1$ line in the Group 1 members. The $^{12}$CO emission exhibits elongated or fan-shaped morphologies originating from the continuum sources. The $^{12}$CO emission can be interpreted as an outflow associated with each source. The apparent lengths at the $3\sigma$ level of these outflows range from $\sim 1000$ to $\sim 6000$ au. S68Nb1 and SMM4A have monopolar outflows, whereas the other four sources have bipolar outflows consisting of blue- and redshifted lobes. Interestingly, the outflows in S68Nc1 and SMM11 are extending roughly parallel to the major axes of their continuum emission (P.A. in Table \ref{tab:info}). The morphology and velocity structures of the outflows in the six sources will be investigated in detail in Section \ref{sec:flow}. 

Figure \ref{fig:gr1}c shows images of the SO line, which is detected at $6\sigma$ levels in all the members of Group 1. SO emission traces various parts of protostellar systems in other star forming regions: for example, a ring between a disk and an envelope in L1527 IRS \citep{ohas14}, a ring in a disk in L1489 IRS \citep{yen14}, a jet in HH212 \citep{lee07}, a disk wind in HH211-mms \citep{lee18}, ambient gas in Barnard 1 \citep{fuen16}. In the case of Group 1, the SO emission traces outflows as seen in Figure \ref{fig:gr1}c. Note that the SO emission from SMM4A and that from SMM4B in Figure \ref{fig:gr1}c(i) are distinct from each other because of the different velocity ranges \citep{aso18}. The SO emission appears to surround the $^{12}$CO outflows or lies in the outer parts of the $^{12}$CO outflow in SMM4A, SMM4B, S68N, and S68Nc1, suggesting that SO traces cavity walls of these $^{12}$CO outflows. However, the SO in S68Nb and SMM11 is stronger in the inner parts of the $^{12}$CO outflows. In addition to the outflows, the SO emission was also detected at the central protostellar positions of SMM4B, S68N, and S68Nc1.

Figure \ref{fig:gr1}d shows images of the C$^{18}$O line. Emission was detected in SMM4B, S68N, S68Nc1, and SMM11. SMM4A shows absorption due to its optically thick continuum emission. It was not detected in S68Nb1, despite extended C$^{18}$O $J=1-0$ emission in IRAM 30 m observations \citep{du-ca10}.

The C$^{18}$O emission in S68Nc1 and SMM11 is elongated in the directions of their $^{12}$CO outflow axes, and velocity gradients of their C$^{18}$O emission are also similar to those of their $^{12}$CO outflows. The C$^{18}$O emission in S68N consists of three components, one centered at the continuum peak position, another in the southwest, and the other in the northeast. The central and southwestern components are much stronger than the northeastern component. The C$^{18}$O emission is elongated in the associated outflow direction at lower contour levels ($<12 \sigma$) around the central component, with its velocity gradient similar to that of the $^{12}$CO outflow. 

The peak {\rm integrated} intensity and the total flux of the C$^{18}$O line in each source were measured inside the $3\sigma$ contour enclosing each source after primary beam correction. The uncertainties of the integrated intensities and the total fluxes are estimated in the same manner as those of the continuum intensities and the continuum flux densities, respectively. When the C$^{18}$O emission is not detected, the $3\sigma$ upper limits were calculated. Those intensities, fluxes, and upper limits are listed in Table \ref{tab:info}. 

\subsection{Group 2}
\begin{figure*}[ht!]
\epsscale{1.1}
\plotone{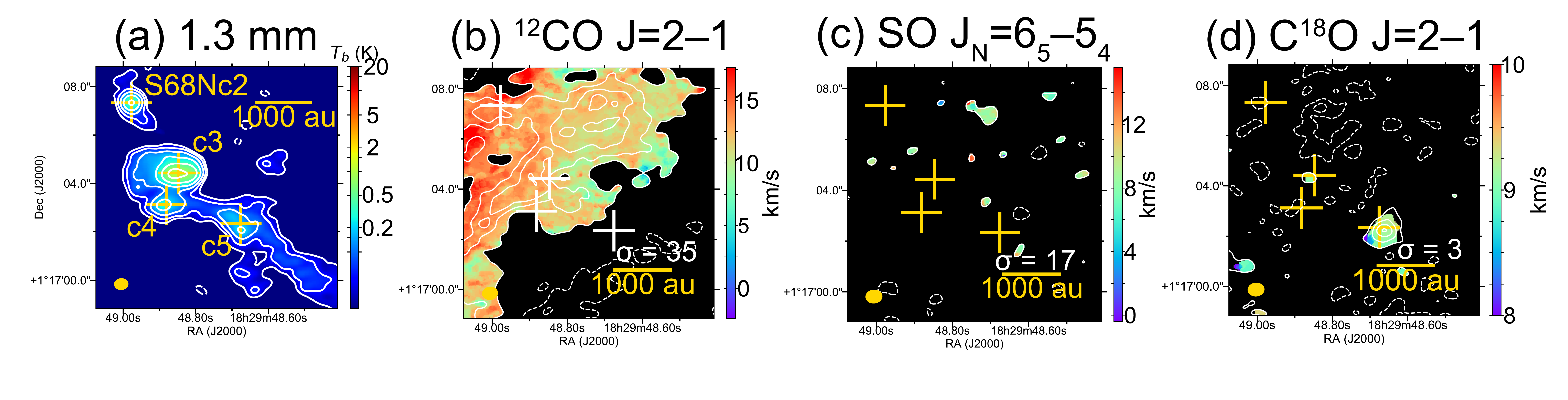}
\caption{Same as Figure \ref{fig:gr1} but for Group 2. The panel sizes are the same as those of columns (a) and (d) in Figure \ref{fig:gr1}.
Contour levels are the same as those of Figure \ref{fig:gr1}. (a) The 1.3 mm continuum. 
(b) The $^{12}$CO $J=2-1$ line. The integrated velocity range is $-41$ - 62 $\kms$.
(c) The SO $J_N=6_5-5_4$ line. The integrated velocity range is $-1$ - 17 $\kms$.
(d) The C$^{18}$O $J=2-1$ line. The integrated velocity range is 7.1 - 7.7 $\kms$.
\label{fig:gr2}}
\end{figure*}

Figure \ref{fig:gr2} shows maps of the 1.3 mm continuum and the $^{12}$CO, SO, and C$^{18}$O lines for the Group 2 members. Gaussian fits and flux measurements were performed in the same manner as those for Group 1, and the results are summarized in Table \ref{tab:info}. For the cases of S68Nc3 and S68Nc4, the fitting was simultaneously done using a double-component 2D Gaussian function. S68Nc3, S68Nc4, S68Nc5, and possibly S68Nc2 as well, are associated with filament $\gtrsim 6000$ au in length. This filamentary structure was excluded in the Gaussian fittings for Group 2.
None of the Group 2 members were identified by Spitzer observations \citep{dunh15}.

Extended $^{12}$CO emission was detected around the Group 2 members, with the exception of S68Nc5. The lack of a systematic velocity gradient suggests that this is ambient gas. The $^{12}$CO emission has a mean velocity of $\sim 12\ \kms$, which is redshifted by $\sim 3\ \kms$ from velocities derived in JCMT, IRAM 30 m, and CARMA observations with 1000s-au spatial resolutions in this region \citep{du-ca10, leek14}. The SO line was not detected toward any of the Group 2 sources. The C$^{18}$O line was clearly detected only in S68Nc5, with marginal detection in S68Nc3. No C$^{18}$O line was detected in the other Group 2 members, although extended C$^{18}$O $J=1-0$ line emission was detected in this region in IRAM 30 m observations \citep{du-ca10}.

\begin{figure*}[ht!]
\epsscale{1.1}
\plotone{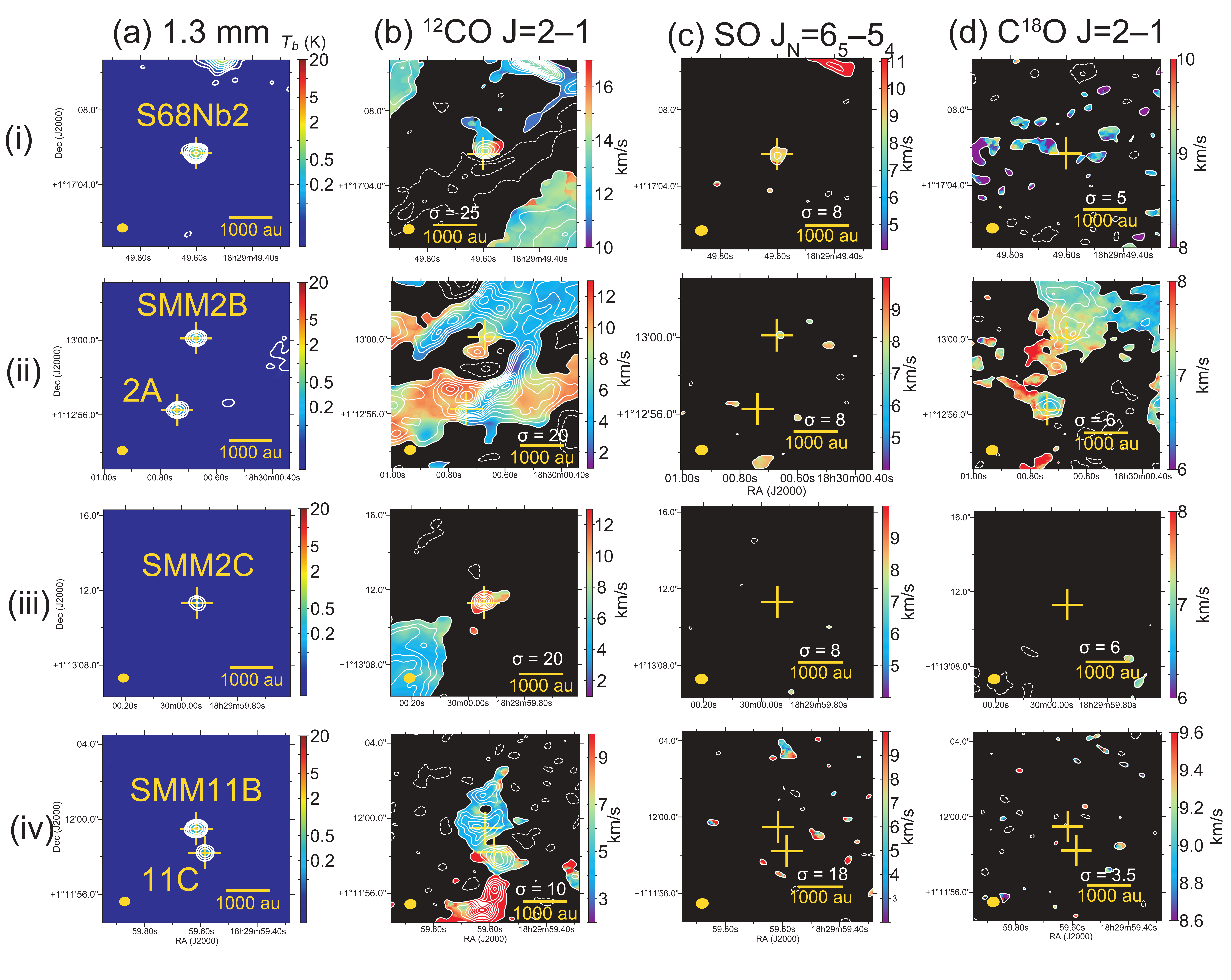}
\caption{Same as Figure \ref{fig:gr1} but for Group 3. The panel sizes are the same as those of columns (a) and (d) in Figure \ref{fig:gr1}. Contour levels are the same as those of Figure \ref{fig:gr1}.
(a) The 1.3 mm continuum. (b) The $^{12}$CO $J=2-1$ line. The integrated velocity ranges are (i) 9 - 18, (ii) $-6$ - 17, (iii) 3 - 5 plus 11 - 17, and (iv) 2 - 5 plus 10 - 21 $\kms$.
(c) The SO $J_N=6_5-5_4$ line. The integrated velocity ranges are  (i) 8 - 10 plus 14 - 16, (ii) 6 - 10, (iii) 6 - 10, and (iv) $-2$ - 18 $\kms$.
(d) The C$^{18}$O $J=2-1$ line. The integrated velocity ranges are (i) 6.3 - 10.1, (ii) 5.3 - 9.6, (iii) 5.3 - 9.6, and (iv) 8.3 - 9.9$\kms$.
\label{fig:gr3}}
\end{figure*}

\subsection{Group 3}
Figure \ref{fig:gr3} shows maps of the 1.3 mm continuum and the $^{12}$CO, SO, and C$^{18}$O lines in the Group 3 members. Gaussian fittings and flux measurements were performed in the same manner as those for Group 1, and the results are summarized in Table \ref{tab:info}. The continuum emission of the Group 3 members arises from regions having deconvolved sizes smaller than half of the beam ($\sim 120$ au). All of the Group 3 members were identified as Class 0 or I protostars by Spitzer observations \citep{dunh15}.

S68Nb2, SMM2C, SMM11B, and SMM11C are associated with compact $^{12}$CO emission within 1000 au of the continuum peaks. 
The $^{12}$CO emission in SMM2A and SMM2B shows more complex structures, although showing a clumpy local peak as well at $\sim 1\arcsec$ southeast of SMM2B. These clumpy $^{12}$CO components of the Group 3 members show spatial offset ($\gtrsim 200$ au) from the continuum peak positions or velocity offset ($\gtrsim 3\ \kms$) from the systemic velocity of Serpens Main (8-9 $\kms$). Hence, their compact $^{12}$CO emission can be interpreted as small outflows. The SO line was detected in S68Nb2 at an LSR velocity of $\sim 9\ \kms$, which is blueshifted by $\sim 6\ \kms$ from the mean velocity of the associated $^{12}$CO emission. Compact C$^{18}$O emission was detected in SMM2A, while extended C$^{18}$O emission was detected in SMM2B. No C$^{18}$O line was detected in the other Group 3 members, although extended C$^{18}$O $J=1-0$ line emission was detected toward regions around all the Group 3 sources with IRAM 30 m observations \citep{du-ca10}.

\begin{deluxetable*}{cccccccc}
\tablecaption{Intensities, fluxes, and sizes of the 1.3 mm sources detected with our ALMA observations. The intensities fluxes, and their uncertainties are primary-beam-corrected, and the upper limits of them are the $3\sigma$ level. $I_{\rm C^{18}O}$ is the integrated intensity at the continuum peak position. $F_{\rm C^{18}O}$ is the total flux of the C$^{18}$O line.
\label{tab:info}}
\tablehead{
\colhead{ } & \colhead{$I_{\rm 1.3mm}^{\rm peak}$} & \colhead{$F_{\rm 1.3mm}$} & \multicolumn{2}{c}{1.3 mm deconvolved FWHM} & \colhead{$I_{\rm C^{18}O}$} & \colhead{$F_{\rm C^{18}O}$} & \colhead{Group in}\\
\colhead{ } & \colhead{$\mJB$} & \colhead{$\mJ$} & \colhead{au $\times$ au (P.A.)} & \colhead{error} & \colhead{$\JB\,\kms$} & \colhead{${\rm Jy}\,\kms$} & \colhead{this paper}
}
\startdata
SMM4A & 196.4$\pm$0.1 & 492$\pm$1 & 322$\times$196 (145$\arcdeg$) & 5$\times$5 (2$\arcdeg$) & absorption & absorption & 1\\
SMM4B & 29.1$\pm$0.1 & 173$\pm$2 & 300$\times$229 (95$\arcdeg$) & 26$\times$19 (16$\arcdeg$ & 0.307$\pm$0.007 & 4.99$\pm$0.13 & 1\\
S68N & 40.4$\pm$0.1 & 208$\pm$2 & 314$\times$217 (38$\arcdeg$) & 19$\times$20 (9$\arcdeg$) & 0.154$\pm$0.006 & 1.59$\pm$0.07 & 1\\
S68Nc1 & 25.6$\pm$0.1 & 114$\pm$2 & 516$\times$238 (87$\arcdeg$) & 25$\times$13 (2$\arcdeg$) & 0.098$\pm$0.008 & 1.01$\pm$0.09 & 1\\
S68Nb1 & 60.2$\pm$0.2 & 108$\pm$2 & 160$\times$143 (86$\arcdeg$) & 6$\times$4 (17$\arcdeg$) & $<$0.027 & $<$0.027 & 1\\
SMM11 & 93.3$\pm$0.1 & 167$\pm$1 & 160$\times$134 (80$\arcdeg$) & 6$\times$5 (10$\arcdeg$) & 0.037$\pm$0.004 & 0.17$\pm$0.02 & 1\\
\hline
S68Nc2 & 5.5$\pm$0.1 & 9.9$\pm$0.7 & 203$\times$122 (19$\arcdeg$) & 25$\times$39 (19$\arcdeg$) & $<$0.009 & $<$0.009 & 2\\
S68Nc3 & 12.3$\pm$0.1 & 52.3$\pm$0.9 & 567$\times$280 (90$\arcdeg$) & 34$\times$18 (3$\arcdeg$) & 0.010$\pm$0.003 & 0.047$\pm$0.003 & 2\\
S68Nc4 & 3.3$\pm$0.1 & 11.4$\pm$0.6 & 497$\times$253 (88$\arcdeg$) & 137$\times$82 (22$\arcdeg$) & $<$0.009 & $<$0.009 & 2\\
S68Nc5 & 2.8$\pm$0.1 & 17.5$\pm$0.9 & 892$\times$471 (63$\arcdeg$) & 99$\times$56 (7$\arcdeg$) & 0.037$\pm$0.003 & 0.062$\pm$0.003 & 2\\
\hline
S68Nb2 & 18.0$\pm$0.2 & 18.0$\pm$0.2 & 30$\times$26 (98$\arcdeg$) & 12$\times$16 (76$\arcdeg$) & $<$0.024 & $<$0.024 & 3\\
SMM2A & 3.8$\pm$0.1 & 4.1$\pm$0.1 & $<$55$\times$41 & -- & 0.027$\pm$0.007 & 0.16$\pm$0.04 & 3\\
SMM2B & 3.5$\pm$0.1 & 3.5$\pm$0.1 & point & -- & 0.028$\pm$0.006 & 1.78$\pm$0.08 & 3\\
SMM2C & 3.1$\pm$0.2 & 3.6$\pm$0.2 &  121$\times$31 (33$\arcdeg$) & 27$\times$58 (25$\arcdeg$) & $<$0.039 & $<$0.039 & 3\\
SMM11B & 25.5$\pm$0.4 & 26.9$\pm$0.4 & 68$\times$22 (95$\arcdeg$) & 3$\times$6 (3$\arcdeg$) & $<$0.042 & $<$0.042 & 3\\
SMM11C & 8.0$\pm$0.4 & 8.6$\pm$0.4 & $<$99$\times$51 & -- & $<$0.042 & $<$0.042 & 3\\
\enddata
\end{deluxetable*}

\subsection{Spectral Energy Distribution of S68Nb1 and S68Nc1} \label{sec:sed}
\begin{figure}[ht!]
\epsscale{1.2}
\plotone{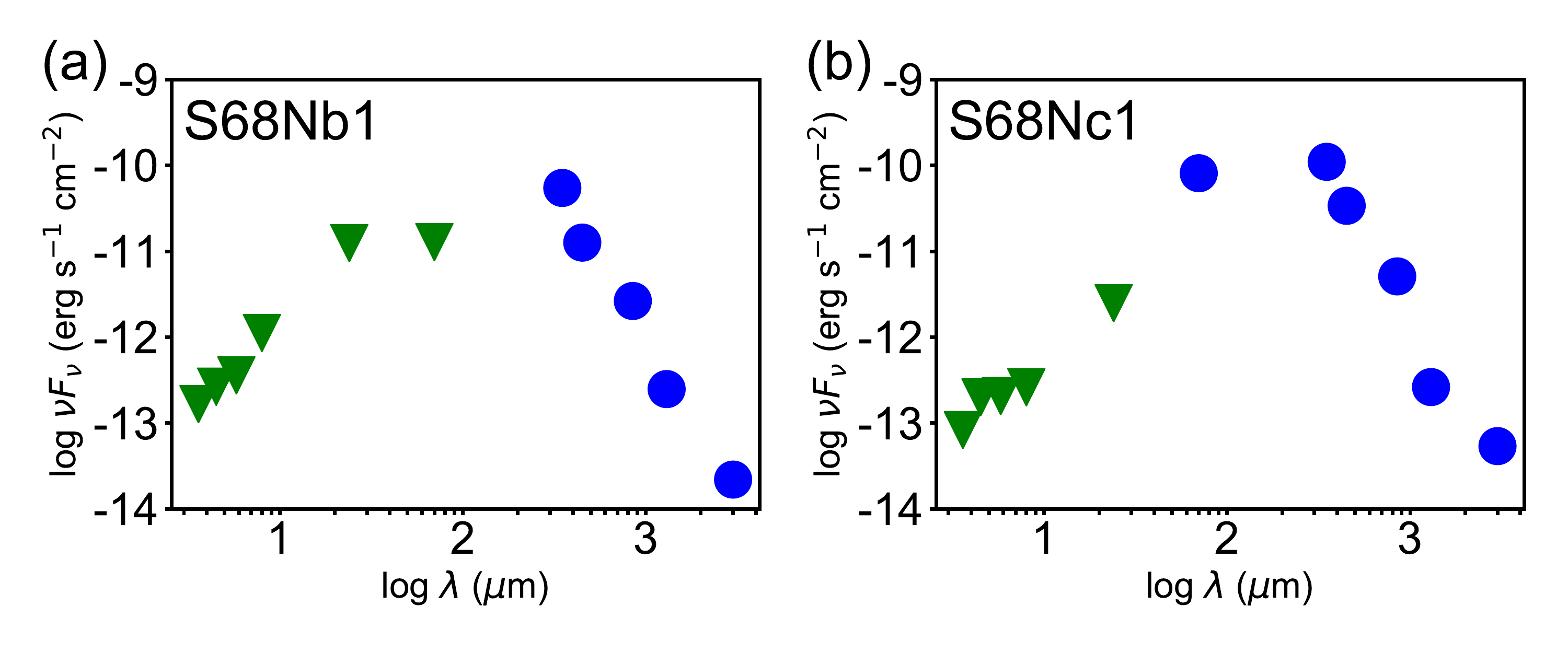}
\caption{Spectral energy distribution of S68Nb1 and S68Nc1. The blue points denote the measured flux densities, while the green points denote detection limits. The uncertainties for the measured values, which are smaller than the symbol size, are listed in Table \ref{tab:sed}.
\label{fig:sed}}
\end{figure}

The presence of the $^{12}$CO outflows (Figure \ref{fig:gr1}b) indicates star formation activities in the Group 1 members. In fact, those except S68Nb1 and S68Nc1 are identified as Class 0 protostars from their SEDs \citep{dunh15, aso17a, aso18}. Hence, the SEDs of S68Nb1 and S68Nc1 are examined to reveal their evolutionary stages in this subsection. Figure \ref{fig:sed} shows the SEDs, bolometric temperature $T_{\rm bol}$, bolometric luminosity $L_{\rm bol}$ , and sub-mm luminosity $L_{\rm smm}$ of S68Nb1 and S68Nc1. The SEDs were constructed from archival data of Spitzer IRAC (3.6, 4.5, 5.8, and 8.0 $\micron$), Spitzer MIPS 24 $\micron$, Herschel PACS 70 $\micron$, CSO SHARK-II 350 $\micron$, and JCMT SCUBA-2 (450 and 850 $\micron$), as well as our ALMA data. In addition, the SEDs also includes the 3 mm flux densities of S68Nb and S68Nc measured with CARMA \citep{leek14}. Our method to derive fluxes and upper limits, and then calculate $T_{\rm bol}$, $L_{\rm bol}$, and $L_{\rm smm}$ is explained by \citet{aso18} in more detail. 
The uncertainties for the detected flux densities were derived in the same way as those for the 1.3 mm flux densities. The flux densities used in the SEDs are listed in Table \ref{tab:sed} with their uncertainties.
Here, we emphasize that the measured fluxes in S68Nc1 and particularly in S68Nb1 include contamination from the neighboring Class I protostar S68Nb2 from mid-infrared to sub-mm wavelengths. Even with such contamination, their upper limits of $T_{\rm bol}\lesssim 60$ K and luminosity ratio $L_{\rm bol} / L_{\rm smm}\lesssim 13$ indicate that both S68Nb1 and S68Nc1 are Class 0 protostars \citep{chen95, andr93}, as the other members of Group 1.

\begin{deluxetable*}{ccccccc}
\tablecaption{Flux densities of S68Nb1 and S68Nc1 in the unit of Jy used in Figure \ref{fig:sed}.
\label{tab:sed}}
\tablehead{
\colhead{ } & \colhead{3.6 $\micron$} & \colhead{4.5 $\micron$} & \colhead{5.8 $\micron$} & \colhead{8.0 $\micron$} & \colhead{24 $\micron$} & \colhead{70 $\micron$}
}
\startdata
S68Nb1 & $<2\times 10^{-4}$ & $<4\times 10^{-4}$ & $<7\times 10^{-4}$ &  $<3\times 10^{-3}$ & $<0.1$ & $<0.3$\\
S68Nc1 & $<1\times 10^{-4}$ & $<3\times 10^{-4}$ & $<4\times 10^{-4}$ & $<7\times 10^{-4}$ & $<0.02$ & $1.9 \pm 0.2$\\
\hline \hline
 & 350 $\micron$ & 450 $\micron$ & 850 $\micron$ & 1300 $\micron$ & 3000 $\micron$ &  \\
\hline
S68Nb1 & $6.4 \pm 0.3$ & $1.9 \pm 0.1$ & $0.75 \pm 0.01$ & $0.108 \pm 0.002$ & $0.022 \pm 0.002$ & \\
S68Nc1 & $12.9 \pm 0.3$ & $5.1 \pm 0.1$ & $1.45 \pm 0.01$ & $0.114 \pm 0.002$ & $0.054 \pm 0.002$
\enddata
\end{deluxetable*}

\section{Analysis} \label{sec:ana}
Sixteen 1.3 mm continuum sources have been detected by our ALMA observations. We have classified them into three groups based on continuum and line characteristics. Here, we derive physical quantities in order to characterize the three groups in more detail.

\subsection{Morphology and Velocity Structures of the Group 1 Outflows} \label{sec:flow}
The Group 1 members show clear outflows in the $^{12}$CO line as seen in Figure \ref{fig:gr1}b. To quantitatively examine these outflows, we first measure the intensity-weighted orientation angles (i.e., P.A.), lengths $r_{\rm flow}$, and opening angles $\theta^{\prime}_{\rm flow}$ projected onto the plane of the sky using $^{12}$CO moment 0 maps. Concretely, these orientation angles and lengths are calculated as the intensity-weighted mean, i.e., $\langle x \rangle \equiv \sum_{ij} I_{ij}x_{ij}/\sum _{ij}I_{ij}$, where $I_{ij}$ and $x_{ij}$ are intensity and the quantity of interest at the pixel $(i, j)$. The uncertainty of the quantity of interest can be calculated from that of the intensity through propagation of uncertainty; the uncertainty of $\langle x \rangle$ is estimated from $(\Delta \langle x \rangle)^2 = (\Delta I)^2\,\sum_{ij}(x_{ij}-\langle x\rangle )^2/(\sum _{ij}I_{ij})^2$, where $\Delta I$ is the uncertainty of the intensity. The derived values are listed in Table \ref{tab:flow}. The outflow opening angles are shown with green lines in Figure \ref{fig:gr1}b as well. In Table \ref{tab:flow}, the outflow direction P.A. is $\langle \theta\rangle$, and the outflow length $r_{\rm flow}$ is $\langle r\rangle $ in the $r$-$\theta$ plane, while the outflow opening angles are calculated as two times intensity-weighted standard deviation $\theta^{\prime}_{\rm flow} = 2\sqrt{\langle \theta ^2\rangle - \langle \theta\rangle ^2}$. Note that, in order to avoid over-estimating the opening angles due to beam convolution, the $^{12}$CO moment 0 maps were deconvolved in the way of CLEAN: finding peaks in a moment 0 map, recording the peaks as ``CLEAN components", subtracting beam (Gaussian) functions from the moment 0 map, and then repeating these down to a $5\sigma$ cutoff. Then, the derived CLEAN components (not the CLEAN components derived from dirty maps in Section \ref{sec:obs}) were adopted as the intensity for weighting. 

The intensity-weighted mean velocities $v_{\rm flow}$ (absolute values) were also measured from the $^{12}$CO data cube using the equation $v_{\rm flow}\equiv |\sum_{k}F_k (v_k - v_{\rm sys})/\sum _k F_k|$, where $F_k$ is the total flux of $^{12}$CO emission associated with each outflow lobe at each velocity channel, $v_k$, and $v_{\rm sys}$ is the systemic velocity. This definition provides an uncertainty of each mean velocity calculated from the uncertainty of the velocity at each channel (i.e., velocity resolution) through propagation of uncertainty, $(\Delta v_{\rm flow})^2 = (\Delta v)^2\,\sum_k F_k^2/(\sum_k F_k)^2$, where $\Delta v$ is the velocity resolution. This requires a systemic velocity of each source. The systemic velocities of SMM4A, SMM4B, and SMM11 were determined from Gaussian fittings to their line profiles in C$^{18}$O $J=2-1$ \citep{aso17a, aso18} to be 7.5, 7.9, and 9.1 $\kms$, respectively. Similarly, the systemic velocities of S68N and S68Nc1 were determined to be 9.3 and 7.6 $\kms$, respectively, by Gaussian fittings to their C$^{18}$O line profiles. The systemic velocity of S68Nb1 was assumed to be the same as that of S68Nc1 since the C$^{18}$O line is not detected in S68Nb1. Dynamical time of the outflows $\tau^{\prime}_{\rm dyn}$, without inclination-correction, was also calculated from $v_{\rm flow}$ and $r_{\rm flow}$. Its uncertainty was calculated from those of $v_{\rm flow}$ and $r_{\rm flow}$ through propagation of uncertainty. The derived $v_{\rm flow}$ and $\tau^{\prime}_{\rm dyn}$ are also listed in Table \ref{tab:flow}.

\begin{figure*}[ht!]
\epsscale{1}
\plotone{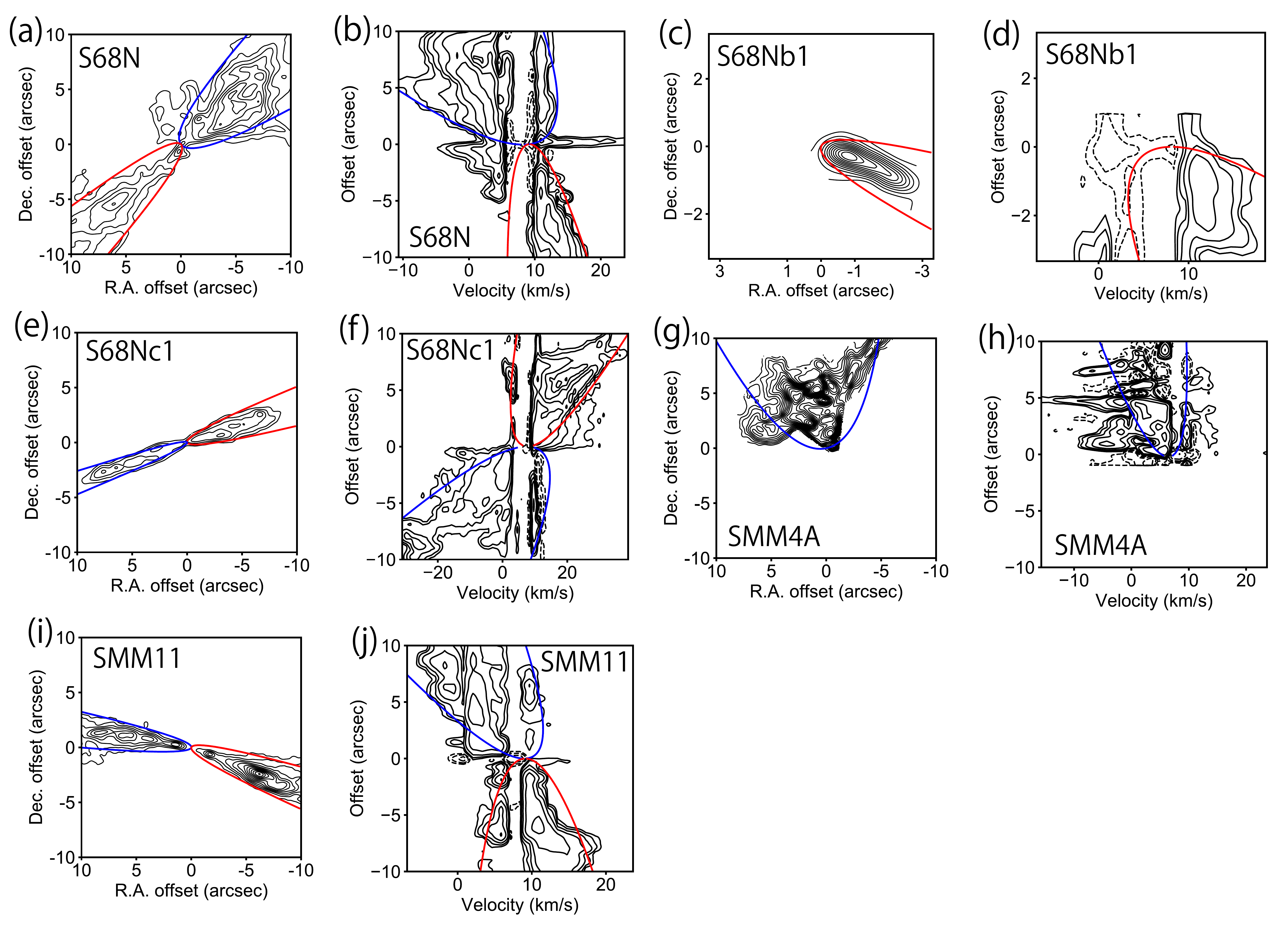}
\caption{Results of the fitting to the $^{12}$CO moment 0 maps and PV diagrams using the wind-driven-shell model. Contour levels of the PV diagrams are 5, 10, 20, 40,$...\times \sigma$, where $1\sigma$ corresponds to $2.6\ \mJB$. The moment 0 maps are the same as the ones in Figure \ref{fig:gr1}b except SMM4A. The integrated velocity range of Figure \ref{fig:gr1}b(i) is selected to include emission from both SMM4A and SMM4B, while the integrated velocity range for SMM4A is $-18$ - 10 $\kms$, and $1\sigma$ of the SMM4A moment 0 map is $50\ \mJB\,\kms$.
\label{fig:fit}}
\end{figure*}

An important uncertainty regarding outflows is their inclination angles $i$. To estimate $i$ of the Group 1 outflows, the $^{12}$CO moment 0 maps and $^{12}$CO position-velocity (PV) diagrams are fitted with the wind-driven-shell model \citep{shu91, lee00}, which adopts a parabolic shape and a radially expanding velocity field. We follow the method described in Appendix A of \citet{yen17}: the $^{12}$CO moment 0 maps (Figure \ref{fig:gr1}b) are used to constrain the morphology, while the velocity and the inclination angle are constrained by fitting PV diagrams along the outflow axes. To simplify the fitting procedure, the intensities at the image pixels with the emission stronger than the 5$\sigma$ level are all replaced with unity, and those weaker than 5$\sigma$ are replaced with zero. The low velocity ranges are filled by emission (i.e., unity) to compensate for the effect of missing flux: 6 - 10 $\kms$ in S68N, 4 - 9$\kms$ in S68Nb1, 4 - 9 $\kms$ in S68Nc1, 7 - 9 $\kms$ in SMM4A, and 7 - 9 $\kms$ in SMM11.
The Markov chain Monte Carlo method is used through an open source, {\it emcee} \citep{fore13}. The log likelihood is $-\chi ^{2} /2$, where $\chi^2$ is the sum of squared differences between the model and the observations (i.e., unity or zero). The uncertainty of the inclination angles are calculated from the 5th and 95th percentiles.
The outflow in SMM4B exhibits the signature of episodic mass ejection \citep{aso18}, which consists of multiple Hubble-law patterns. Our single shell model is not appropriate to fit such a complex velocity structure. For this reason, SMM4B is excluded for this fitting, and $i$ and its uncertainty of the SMM4B outflow are adopted from \citet{aso18}, which was estimated by assuming equal momentum ejection between the blue- and redshifted lobes of the SMM4B outflow. Figure \ref{fig:fit} shows the best-fit model curves overlapped with the moment 0 maps and PV diagrams in the $^{12}$CO line. The best-fit $i$ is listed in Table \ref{tab:flow}. 

\begin{figure}[ht!]
\epsscale{1}
\plotone{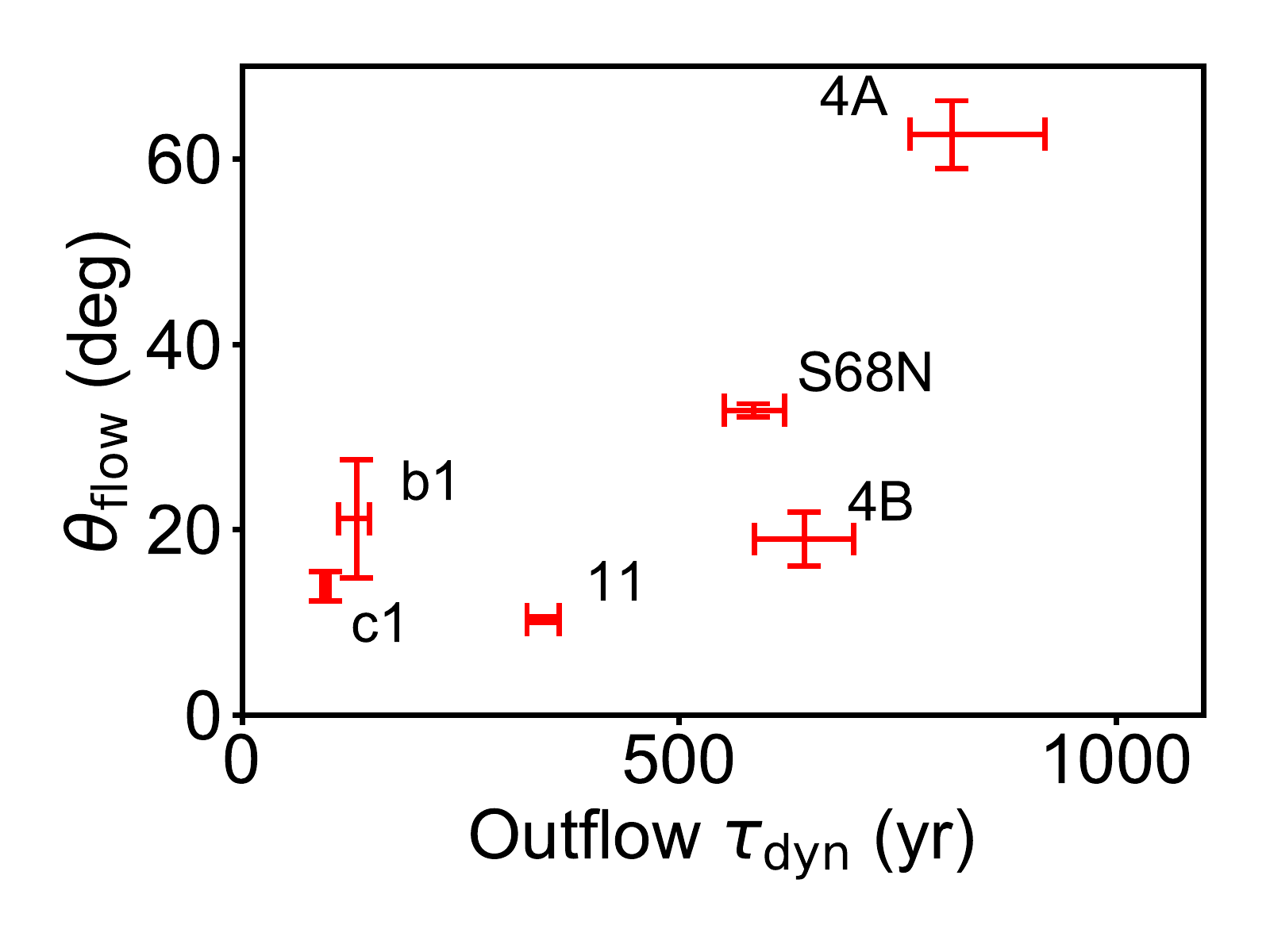}
\caption{Dynamical time $\tau _{\rm dyn}$ and opening angles $\theta _{\rm flow}$ of the Group 1 outflows. Both abscissa and ordinate values are inclination-corrected using the inclination angles in Table \ref{tab:flow}. Mean values are plotted for the outflows with blue- and redshifted lobes. 
\label{fig:flow}}
\end{figure}

The intensity-weighted values are corrected by using the inclination angles derived from the fitting. Figure \ref{fig:flow} shows the relation between inclination-corrected $\tau _{\rm flow}$ and $\theta _{\rm flow}$. The inclination-corrected values $\tau_{\rm dyn}$ and $\theta _{\rm flow}$ are defined as $\tau_{\rm dyn} = \tau ^{\prime}_{\rm dyn} / \tan i$ and $\tan (\theta _{\rm flow}/2) = \sin i\, \tan (\theta^{\prime}_{\rm flow}/2)$ using the quantities in Table \ref{tab:flow}. 
Figure \ref{fig:flow} shows the mean values for the outflows having both blue- and redshifted lobes. The errors in this figure are calculated from those of $\tau^{\prime}_{\rm dyn}$, $\theta ^{\prime}_{\rm flow}$, and $i$ through propagation of uncertainty. Although the number of the sampling points is small, the points in Figure \ref{fig:flow} seem to show a hint of trend.

\begin{deluxetable*}{rrrrrrr}
\tablecaption{Properties of the $^{12}$CO outflows in Group 1. The inclination angle $i$ is measured from the line-of-sight ($i=0\arcdeg$ is pole-on). The sixth column is dynamical time calculated as $\tau ^{\prime}_{\rm dyn}=r_{\rm flow}/v_{\rm flow}$. $r_{\rm flow}$, $\theta ^{\prime} _{\rm flow} $, $v_{\rm flow}$, and $\tau ^{\prime}_{\rm dyn}$ in this table are not inclination-corrected.
\label{tab:flow}}
\tablehead{
\colhead{ } & \colhead{P.A.} & \colhead{$r_{\rm flow}$} & \colhead{$\theta ^{\prime} _{\rm flow}$} & \colhead{$v_{\rm blow}$} & \colhead{$\tau ^{\prime}_{\rm dyn}$} & \colhead{$i$} \\
\colhead{ } & \colhead{(deg)} & \colhead{(au)} & \colhead{(deg)} & \colhead{($\kms$)} & \colhead{(year)} & \colhead{(deg)}
}
\startdata
SMM4A (blue) & 14.1$\pm$1.7 & 2030$\pm$40 & 67.8$\pm$3.7 & 5.5$\pm$0.3 & 1750$\pm$100 & 65.1$^{+0.3}_{-2.6}$ \\
SMM4B (blue) & 6.1$\pm$0.9 & 1900$\pm$60 & 23.9$\pm$3.9 & 13.4$\pm$0.2 & 670$\pm$20 & 36$^{+3}_{-3}$ \\
- (red) & 146.0$\pm$1.2 & 2000$\pm$110 & 25.4$\pm$5.4 & 9.5$\pm$0.2 & 1000$\pm$60 & 70$^{+2}_{-2}$ \\
S68N (blue) & -51.0$\pm$0.4 & 2940$\pm$20 & 35.6$\pm$1.4 & 5.8$\pm$0.4 & 2400$\pm$170 & 72.0$^{+0.1}_{-0.3}$ \\
- (red) & 133.0$\pm$0.5 & 4280$\pm$60 & 32.1$\pm$0.2 & 6.0$\pm$0.5 & 3380$\pm$290 & 83.4$^{+0.1}_{-0.5}$ \\
S68Nc1 (blue) & 109.6$\pm$0.5 & 2250$\pm$90 & 11.0$\pm$1.3 & 14.7$\pm$0.3 & 730$\pm$30 & 84.8$^{+0.1}_{-0.1}$ \\
- (red) & -71.8$\pm$0.6 & 1780$\pm$70 & 16.9$\pm$3.0 & 9.2$\pm$0.4 & 920$\pm$50 & 82.3$^{+0.1}_{-0.1}$ \\
S68Nb1 (red) & -111.9$\pm$1.3 & 640$\pm$20 & 21.8$\pm$6.6 & 5.6$\pm$0.6 & 540$\pm$60 & 76.4$^{+1.5}_{-0.5}$ \\
SMM11 (blue) & 82.4$\pm$0.2 & 4020$\pm$60 & 11.1$\pm$0.5 & 5.9$\pm$0.3 & 3230$\pm$170 & 81.0$^{+0.1}_{-0.3}$ \\
- (red) & -109.1$\pm$0.2 & 4430$\pm$50 & 9.6$\pm$0.3 & 4.2$\pm$0.4 & 5000$\pm$480 & 88.0$^{+0.1}_{-0.2}$ \\
\enddata
\end{deluxetable*}

\subsection{Jeans Stability of the Group 2 Members} \label{sec:jeans}
The Group 2 members are located in the filamentary structure in the S68Nc region (Figure \ref{fig:gr2}a), and have 1.3 mm continuum fluxes several times lower than the Group 1 protostars (Table \ref{tab:info}). None of them shows a signature of a $^{12}$CO outflow or a Spitzer source \citep{dunh15} which
suggests that Group 2 members are starless. The gravitational stability of the Group 2 cores is assessed through a Jeans analysis. Central densities are calculated from the 1.3 mm continuum intensities in Table \ref{tab:info}. First, the peak intensity is converted to a column density of molecular hydrogen within the $\sim 0\farcs 5$ ($\sim$ 210 au) beam. We adopt a dust opacity coefficient of $\kappa (850\ \micron) = 0.035\ {\rm cm}^2\,{\rm g}^{-1}$ \citep{an.wi05}, an opacity index, $\beta=1$, and a temperature of 10 K because the Group 2 members have no central heating sources. Secondly, the H$_2$ column density is converted to the central number density $n_{\rm H_2}$ by assuming a Gaussian density profile along the line of sight with a FWHM of the geometrical mean of the deconvolved sizes along the major- and minor-axes listed in Table \ref{tab:info}. Then, the Jeans length $\lambda _{\rm Jeans}$ is calculated as $\lambda _J = c_s \sqrt{\pi / (G \mu m_{\rm H} n_{\rm H_2})}$, where $c_s$, $G$, $\mu$, $m_{\rm H}$ are the sound speed at 10 K, the gravitational constant, the mean molecular weight ($=2.37$), and the mass of atomic hydrogen.

\begin{figure}[ht!]
\epsscale{1}
\plotone{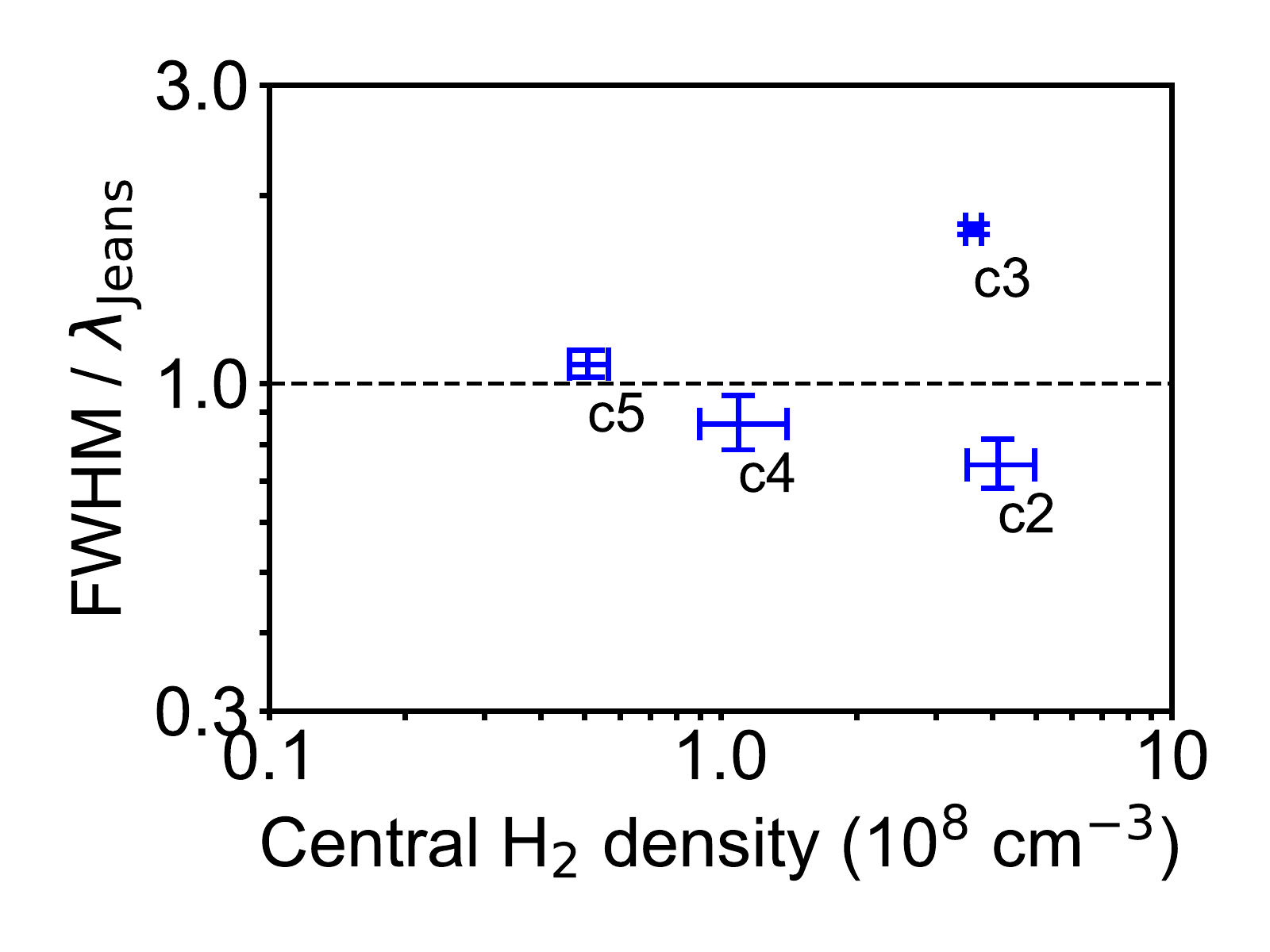}
\caption{Relation between the central densities and the ratios between the deconvolved FWHM (geometrical mean along the major and minor axes) and the Jeans lengths for Group 2. These quantities are calculated from the 1.3 mm intensities and the deconvolved FWHMs in Table \ref{tab:info}. The temperature is assumed to be $T=10$ K because the Group 2 members are not associated with protostars that can be heating sources. The values and their errors in this figures also assume a dust opacity of $\beta=1$. Different $T$ and $\beta$ shift all the values together upward or downward.
\label{fig:dens}}
\end{figure}

Figure \ref{fig:dens} shows the relation between $n_{\rm H_2}$ and the ratio FWHM$/\lambda _{\rm Jeans}$ for Group 2. The errors in this figures are calculated from those of the peak intensity, major-axis, and minor axis through propagation of uncertainty. In addition, if the dust opacity index $\beta$ is close to the interstellar value $\sim2$, the central H$_2$ density is $\sim50\%$ higher, and the ratio FWHM$/\lambda _{\rm Jeans}$ is $\sim 20\%$ higher. If the dust is more evolved ($\beta \sim 0$), these quantities are lower by the same factor. If the temperature is relatively high as prestellar sources, $\sim 15$\,K, the density is two times lower, and the ratio is $\sim 40\%$ lower. The absolute flux error $\sim 10$\% also causes uncertainties of $\sim 10$\% and $\sim 3$\% for the density and the ratio, respectively. The ratios are around unity within the total uncertainty of a few 10\%, implying that the Group 2 sources are marginally Jeans stable/unstable. The central densities of the Group 2 members are typically $\sim 10^8\ {\rm cm}^{-3}$, which is similar to that of a prestellar Bonner-Ebert sphere on the spatial scale of our angular resolution, $\sim 200$ au \citep[e.g.,][]{aika08}. For these reasons, we interpret the Group 2 members as prestellar sources. 

\subsection{C$^{18}$O Abundance}
\begin{figure}[ht!]
\epsscale{0.9}
\plotone{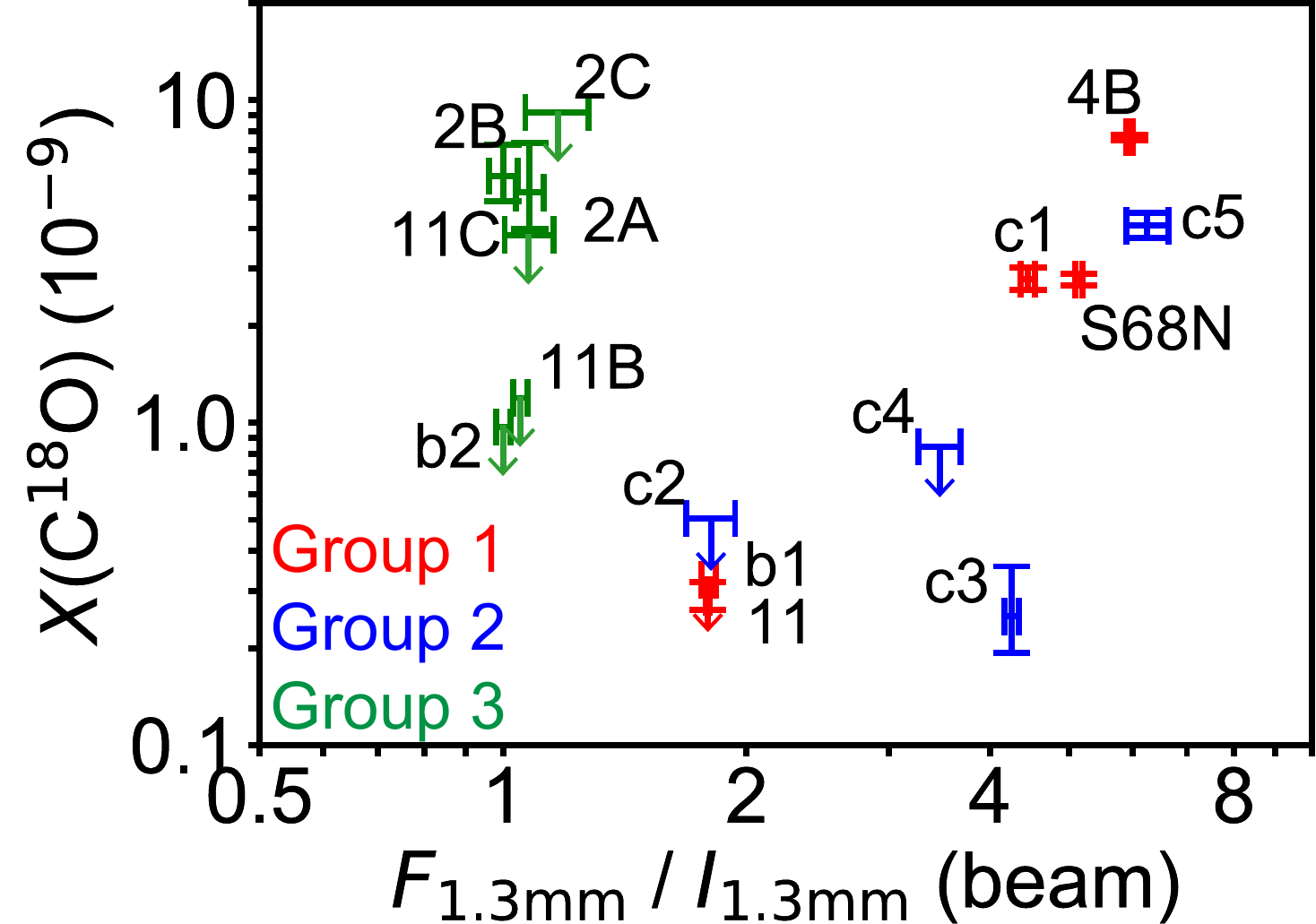}
\caption{Variation of the central concentration degree of the 1.3 mm continuum emission and the C$^{18}$O fractional abundance. The C$^{18}$O abundance is calculated from the 1.3 mm intensity and the C$^{18}$O integrated intensity at the continuum peak position by assuming optically thin emission. The temperature is assumed to be 20 K for the protostellar sources (Group 1 and 3) \citep{leek14} and 10 K for the prestellar sources (Group 2). SMM4A is not plotted because the C$^{18}$O line is detected as absorption against strong continuum emission, preventing us from calculating the C$^{18}$O abundance.
\label{fig:c18o}}
\end{figure}

The fractional abundance of C$^{18}$O relative to H$_2$ is calculated from the 1.3 mm continuum intensity and the C$^{18}$O integrated intensity at the continuum peak position, assuming both are optically thin. The temperature is assumed to be 20 K for the protostellar sources (Group 1 and 3) \citep{leek14} and 10 K for the prestellar sources (Group 2), and local thermodynamic equilibrium is assumed. SMM4A is excluded from this estimation because the C$^{18}$O line is detected as absorption in this source due to its optically thick continuum emission. Figure \ref{fig:c18o} shows the derived C$^{18}$O abundance. The abscissa, central concentration degree, is plotted merely to distinguish the sources. The errors in this figure are calculated from those of the continuum peak intensity, continuum flux density, and C$^{18}$O integrated intensities through propagation of uncertainty. 
The optical depths in SMM11 and S68Nb1 are estimated to be $\sim 0.8$ and $\sim 0.4$, and lower for the other sources.
When these optical depths are considered, the estimated $X({\rm C}^{18}{\rm O})$ is higher by a factor of $(e^{\tau _{c}} -1)/\tau _{c}$, $\sim 1.5$ and $\sim 1.2$ for SMM11 and S68Nb1 respectively, where $\tau _{c}$ is the optical depth of continuum emission. These factors, however, do not change overall trends of data points in Figure \ref{fig:c18o}.

Compared to the C$^{18}$O abundance in Serpens Main of 6-$9\times10^{-8}$
at $\sim 15\arcsec$ ($\sim 6400$ au) scales based on JCMT and IRAM 30 m observations \citep{du-ca10},
the calculated abundances of the pre- and proto-stellar cores are 10-100 times lower. Figure \ref{fig:c18o} shows that the C$^{18}$O abundance varies by an order of magnitude even in each of Group 1, 2, and 3.
This variation will be discussed in the context of evolution in Section \ref{sec:abu}.

\section{Discussion} \label{sec:dis}
\subsection{Evolutionary Trends within the Class 0 Stage} \label{sec:evo}
All of the Group 1 members are classified as Class 0 sources. Nevertheless, our previous study \citep{aso18} on three of them, SMM11, SMM4A, and SMM4B revealed that each have distinct evolutionary characteristics. Specifically, SMM4A has the largest disk and the widest outflow opening angle; SMM4B has an unresolved disk, whereas SMM11 does not show presence of a disk in the $uv$ domain; C$^{18}$O is more abundant in SMM4 than in SMM11. 
This picture is consistent with the classical scenario of star formation \citep[e.g.,][]{tere84, basu98}, in which disks grow steadily to the sizes of $r \sim 100$ au, as in the case of SMM4A. Our discovery of Group 3 sources with small disks, however, requires that the classical disk growth scenario should be modified to consider a range of final disk radii Currently none of the Group 1 sources except SMM4A has a large disk. These members, therefore, could evolve to become Group 3-like objects (Section \ref{sec:gr3}). In order to avoid the bias towards the classical disk growth scenario, we exclude SMM4A in this subsection.
The relation between SMM4A and the other Group 1 members will be discussed in Section \ref{sec:gr3}. We use a simple scoring system to assess evolutionary trends in the other Group 1 members here, and summarize our conclusions in Table \ref{tab:evo}.

\subsubsection{Outflows}
Previous theoretical \citep{ma.ho13} and observational \citep{ar.sa06} studies suggest outflow widening over the course of protostellar evolution. The relation between the dynamical time, $\tau _{\rm dyn}$, and the opening angle, $\theta _{\rm flow}$, of the Group 1 outflows is consistent with this picture.
However, the difference of $\theta_{\rm flow}$ is less clear than that of $\tau_{\rm dyn}$. Thus, using only $\tau _{\rm dyn}$, we score 1 point to SMM4B and S68N, having $\tau _{\rm dyn} > 500$ yr, and 0 point to the rest, having $\tau _{\rm dyn}<500$ yr.

The spatial distribution of the SO emission shown in Figure \ref{fig:gr1}c can also be explained in an evolutionary sequence as follows. S68Nb1 and SMM11 show jet-like, collimated SO emission in inner parts of the associated $^{12}$CO outflows, i.e., in the vicinity of the outflow axes. In contrast, the SO emission occurs in outer parts of their $^{12}$CO outflows and traces the walls of the evacuated cavities.
Jet-like, collimated components are reported only in young Class 0 protostars: for example, HH212 shows SO emission tracing such a jet-like, collimated component \citep{lee07}. Furthermore, S68Nb1 and SMM11 do not show SO emission at their protostellar positions, unlike the others. This difference can be explained by rising temperature: warm regions where SO is in the gas phase ($>50$ K) is smaller in younger phases, and thus the SO emission is more diluted by our beam size ($\sim 200$ au). Even if SO is enhanced by accretion onto the disks, the shocked regions, i.e., the outskirts of the disks are smaller at earlier times, and thus such SO emission would be weaker. For these reasons, we score 1 point to SMM4B, S68N, and S68Nc1, showing SO emission in the outer parts of the outflows and at the protostellar positions, and 0 point to the rest, showing jet-like SO emission and no SO emission at the prostellar positions.  


\subsubsection{C$^{18}$O Abundance} \label{sec:abu}
The C$^{18}$O abundances of SMM11 and S68Nb1 are lower than those of S68Nc1, S68N, and SMM4B (Figure \ref{fig:c18o}), suggesting that the former two are younger than the latter three from the viewpoint of CO desorption \citep{aika12}. \citet{ar.sa06} proposed that C$^{18}$O emission is enhanced by the heating from young outflows, similar to S68Nc1 here, which has C$^{18}$O emission along its outflow. The C$^{18}$O emission becomes stronger in envelopes later, as in S68N and SMM4B. We score 2 points to SMM4B and S68N, showing desorption ($X({\rm C}^{18}{\rm O})>10^{-9}$) in their envelopes, 1 point to S68Nc1, showing desorption in its outflow, and 0 point to the rest, showing freeze-out ($X({\rm C}^{18}{\rm O})<10^{-9}$).

The variation in the C$^{18}$O abundances of sources in Group 2 and 3 can also be explained by evolution.
Among the Group 2 members, S68Nc5 has the lowest central H$_2$ density and highest $X({\rm C}^{18}{\rm O})$.
The temperature is expected to be higher in lower density cores due to greater cosmic ray heating, and
theoretical models show that the C$^{18}$O abundance decreases with declining temperature as a prestellar source evolves to a more centrally concentrated state \citep{aika08}.
These considerations may explain the variation of the C$^{18}$O abundance seen in Figure \ref{fig:c18o} among the Group 2 members and suggest that SMM11 and S68Nb1 are closer to the prestellar phase than the other Group 1 members.
The Class 0 members of Group 3, SMM2A and SMM2B, have similar C$^{18}$O abundances to the highest values in Group 1. The other Group 3 members are more evolved, Class I, sources with more dissipated envelopes and have lower C$^{18}$O abundances.
This variation of the C$^{18}$O abundance among the Group 3 members may be explained by one, or both, of the following mechanisms.
\citet{hars15} show that an envelope can heat the inner disk and suppress volatile freeze-out.
In addition, the C$^{18}$O abundance will steadily decrease as the disk evolves from Class 0 to I through the chemical sink mechanism \citep{fu.ai14}.

\subsubsection{Continuum Morphology} \label{sec:mor}
\begin{figure}[ht!]
\epsscale{1.2}
\plotone{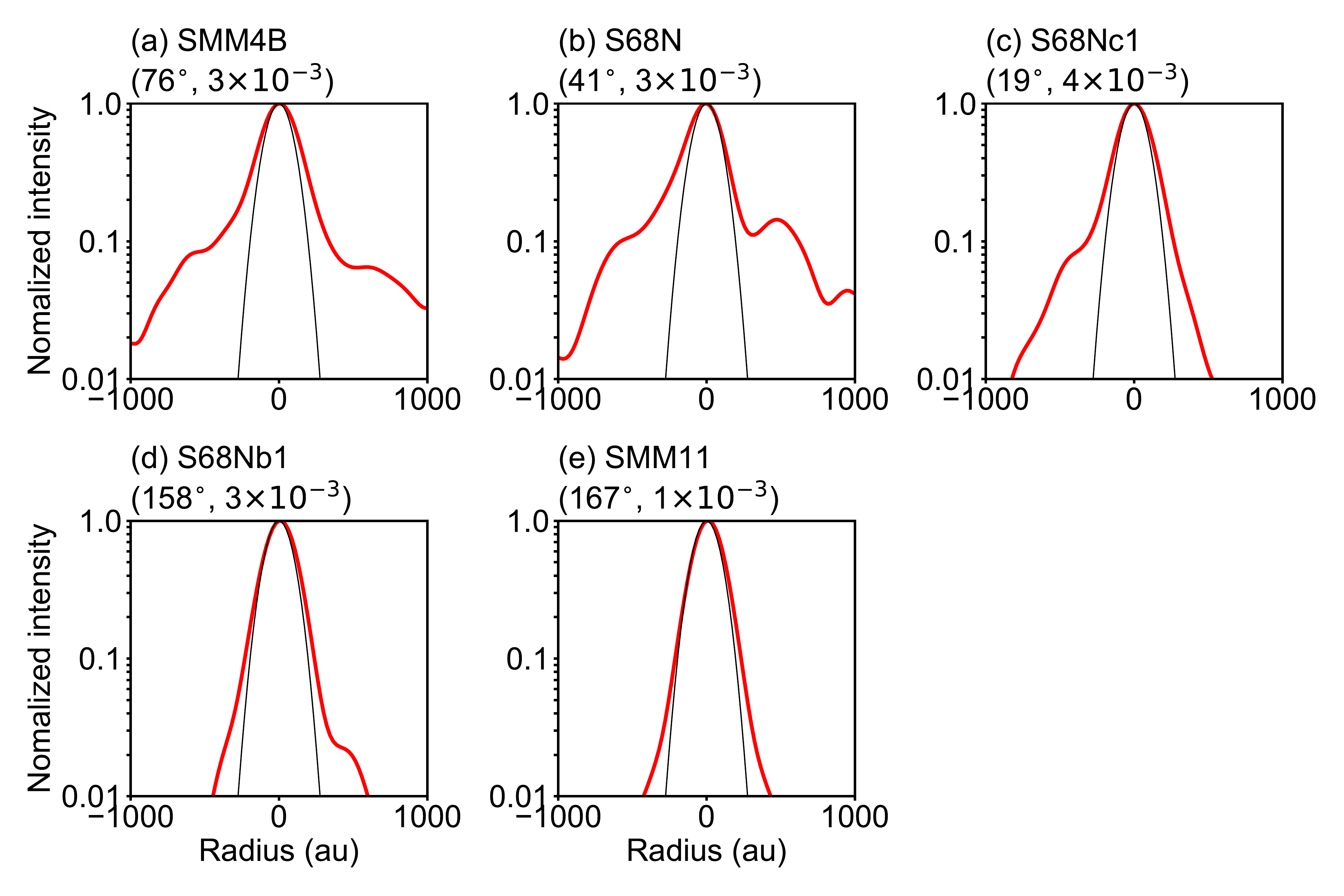}
\caption{Continuum intensity profiles of the Group 1 members except SMM4A in the directions perpendicular to their outflows across their protostellar positions (red thicker curves). Position angles of the negative position offset (abscissa) and noise levels normalized by the peak intensities are denoted below the source names. The black thinner curves denote our beam size ($0\farcs5$). S68N, SMM4B, and S68Nc1 clearly show extended components as well as central compact components. Although S68Nb1 and SMM11 show single components, their emission is significantly wider than the beam size.
\label{fig:prof}}
\end{figure}

Figure \ref{fig:prof} shows that the continuum intensity profiles perpendicular to the outflows across the protostellar positions. The profiles of SMM4B, S68N, and S68Nc1 includes extended components and central compact components significantly stronger than the extended components. We interpret the extended and compact components as envelopes and unresolved disks, respectively.
The extended component of S68Nc1 is elongated along its outflow direction or tracing a cavity wall of the outflow (Figure \ref{fig:gr1}b(iii)). S68Nb1 and SMM11 do not show such compact$+$extended components, although their emission is significantly more extended than the expected emission from a central point source. Their single-peak structures suggest that any central disks are too small to be detected. We score 1 point to SMM4B, S68N, and S68Nc1, showing unresolved disks, and 0 point to the rest, showing no detectable disks. 

S68Nb1 and SMM11 show single compact continuum emission. Furthermore, their continuum emission is close to circular (aspect ratio $\sim 1.2$), and their deconvolved major axes (P.A. in Table \ref{tab:info}) are in similar directions to their outflow axes. S68Nc1 also shows continuum emission elongated along its outflow axis. Such morphologies suggest spherical envelopes and outflows, rather than flattened envelopes and disks \citep{aso17a}. We thus score 1 point to SMM4B and S68N and 0 point to the rest, based on whether they lack or exhibit
continuum elongations along their outflows.

The continuum emission in SMM11 is elongated along the north-south direction at low contour levels in Figure \ref{fig:gr1}a(v) (it is more clearly seen in the zoom out version in Figure 1a of \citet{aso17a}). This elongation reflects the structure of the filament passing SMM11 on a sub-pc scale \citep{leek14}. Such filamentary structures are interpreted as gas accretion streams in other young protostars, such as L1157 \citep{loon07} and L1521F \citep{toku17}. The elongated component of the continuum emission in SMM11 is thus consistent with the conclusion that SMM11 is relatively young in Group 1. A similar filamentary structure is seen more clearly around the prestellar sources (Group 2), i.e., even younger sources.

\begin{deluxetable*}{cccccc|l}
\tablecaption{Evolutionary scores of the Group 1 members except for SMM4A.
\label{tab:evo}}
\tablehead{
\colhead{ } & \colhead{$\tau _{\rm dyn}$ of} & \colhead{distribution of} & \colhead{C$^{18}$O} & \colhead{Unresolved disk} & \colhead{P.A.$_{\rm flow}-$P.A.$_{\rm cont}$} & \colhead{Total} \\
\colhead{ } & \colhead{$^{12}$CO outflow} & \colhead{SO emission} & \colhead{desorption} & \colhead{at 1.3 mm} & \colhead{} & \colhead{score}
}
\startdata
SMM4B   & 1 ($>500$ yr) & 1 (outflow, center) & 2 ($>10^{-9}$ in envelope) & 1 (detected) & 1 ($\gtrsim 70\arcdeg$) & 6 (evolved) \\
S68N    & 1 ($>500$ yr) & 1 (outflow, center) & 2 ($>10^{-9}$ in envelope)& 1 (detected)& 1 ($\gtrsim 70\arcdeg$) & 6 \\
S68Nc1  & 0 ($<500$ yr) & 1 (outflow, center) & 1 ($>10^{-9}$ in outflow) & 1 (detected)& 0 ($\lesssim 20\arcdeg$) & 3 \\
S68Nb1  & 0 ($<500$ yr) & 0 (jet, no center) & 0 ($<10^{-9}$) & 0 (no) & 0 ($\lesssim 20\arcdeg$) & 0 \\
SMM11   & 0 ($<500$ yr) & 0 (jet, no center) & 0 ($<10^{-9}$) & 0 (no) & 0 ($\lesssim 20\arcdeg$) & 0 (young) \\
\enddata
\end{deluxetable*}

One may wonder why the young members, S68Nc1, S68Nb1, and SMM11 show smaller envelopes in the 1.3 mm continuum than the more evolved members, S68N and SMM4B. 
Similar results are found in the Barnard 1 region between a typical Class 0 protostar, B1-c, and young Class 0 protostars, B1-bS and B1-bN \citep{hi.li14}. B1-c shows 0.9 mm continuum emission extending over $\sim 4\arcsec$ \citep{cox18}, whereas B1-bS and B1-bN show 0.9 mm continuum emission extending over $\sim 0\farcs6$ \citep{geri17}. The apparent sizes of the continuum emission are related to density and temperature distributions. The density at a given radius rises when the mass-supplying radius expands outward because outer regions have more volume and thus more mass. The temperature distribution depends on the luminosity of the central protostar. For example, S68N has $L_{\rm bol}=14\ \Ls$ \citep{dunh15}, whereas the younger members have $L_{\rm bol}\lesssim 2\ \Ls$. The $\sim 7$ times larger $L_{\rm bol}$ can cause a $\sqrt{7}= 2.6$ times larger iso-temperature radius \citep[e.g.,][]{go.kw1974}. In addition, the more extended members, SMM4B, S68N, and S68Nc1, show lower 1.3 mm peak intensities, implying lower densities, than the more centrally concentrated members, S68Nb1 and SMM11. Due to this density difference, the central star could heat the envelope more easily in the extended three members than in the compact two members. Although it is not clear only from the present data whether these two mechanisms are due to protostellar evolution or intrinsic differences, the combination of these mechanisms would explain the difference of the apparent envelope sizes in the 1.3 mm continuum among the Group 1 members.

\subsection{Group 3 and Disk-Size Diversity} \label{sec:gr3}
Six 1.3 mm continuum sources are categorized into Group 3 in our sample. Spitzer observations \citep{dunh15} identified four members (SMMb2, SMM2C, SMM11B, and SMM11C) as Class I protostars and two members (SMM2A and SMM2B) as Class 0 protostars (Table \ref{tab:info}). Their bolometric luminosities $L_{\rm bol}\gtrsim 1\ \Ls$ indicate that the central objects are massive enough to be classified as protostars rather than proto-brown dwarfs. Their total flux densities of the 1.3 mm continuum emission range from 27 down to 3 $\mJ$. This range corresponds to the gas-mass range from 0.045 down to 0.006 $\Ms$, where $\kappa(1.3\ {\rm mm}) = 0.023\ {\rm cm^2\,g^{-1}}$ and $T_{\rm dust} = 20$ K are assumed. This is typically $>10$ times smaller than those of Group 1 (0.2 - 0.8 $\Ms$). The Group 3 members also show compact, faint outflows in the $^{12}$CO line. These results, i.e., Spitzer identification, the deficiency of the envelope materials, and the faint outflows, can be explained by a final phase of mass accretion, where most of the envelope materials are being exhausted. The lack of envelope causes outflows to be faint in molecular lines, while making the central protostar bright even at near- and mid-IR wavelengths.

The envelope dissipation around the Group 3 members suggests that Group 3 disks will not grow further from their present size, $r\lesssim 15$-$60$ au (Table \ref{tab:info}). In this sense, Group 3 may represent the precursors of small T Tauri disks with tens-au radii, identified in recent observational studies \citep{ciez19, na.be18}. In particular, SMM2A and SMM2B show $r\lesssim 30$ au at the Class 0 stage, whereas a large disk ($r\sim 240$ au) is identified from a continuum visibility analysis in the Group 1 member SMM4A at the same stage \citep{aso18}. These differences imply a diverse evolution from the Class 0 phase, which may be due to local conditions such as multiplicity or magnetic field configurations as discussed in Section \ref{sec:ori3} in more detail, even in the same star forming region. This diversity in protostellar initial conditions may produce a disk-size diversity over radii of tens to hundreds au in the T Tauri phase \citep{na.be18}. 


Previous observations also identified Group-3-like objects in other star forming regions at Class 0 and I stages. For example, SM1-A (Class 0) and Source-X (Class 0) in Oph A \citep{kawa18}, B1-bW (Class I) in Barnard 1 \citep{hi.li14}, L1448C(S) (Class I) in Perseus \citep{hira10, tobi15}, and MMS-1 (Class I) in MC27/L1521F \citep{toku17, bour06} show small ($\lesssim 100$ au) and faint (i.e., low-mass $\lesssim 0.05\ \Ms$) or no mm/sub-mm continuum emission and clumpy $^{12}$CO outflows.
This suggests a similar diversity of protostellar conditions in other star-forming regions as well as Serpens Main.

\subsection{Possible Formation Mechanisms of Group 3}
\label{sec:ori3}
Theoretical studies provide potential mechanisms for the formation of the Group 3 members. A relatively simple mechanism is star formation in a very low mass ($\sim 0.1\ \Ms$) core \citep{tomi10}
In the case of the binary formation, most of the initial angular momentum is converted to the orbital motion, leaving only small angular momentum to form circumstellar disks, as found in a binary system, L1551 NE \citep{taka17}; this could be the case for the closest pair of Group 3 members SMM11B and SMM11C with a seperation of $\sim 600$ au. Similarly ejection from multiple systems can make disk masses and disk sizes smaller \citep{bate18}. Another potential mechanism is the Hall effect in MHD; magnetic fields parallel to the initial angular momentum vector suppress disk growth, while anti-parallel combination enhances disk growth, resulting in a bimodal distribution of disk sizes \citep{tsuk17}. In addition to the Hall effect, a low initial ratio of rotational to gravitational energies or low diffusivity in non-ideal MHD can also enhance magnetic breaking, and thus produce smaller disks.

Observational studies provide other potential mechanisms as well. Brown dwarfs tend to have less massive disks than typical low mass stars \citep[e.g.,][]{scho06}, and such disks could appear smaller at a given sensitivity. However, the Group 3 members are not proto-brown dwarfs, and their bolometric luminosities are similar or higher than those of the Group 1 members. The Group-3-like source L1448C(S) is found to be impacted by an outflow from a neighboring protostar, L1448C(N) \citep{hira10}, which could blow off the envelope around L1448C(S), and terminate mass accretion forcibly. The Group 3 members in Serpens Main may be in the same situation because large scale outflows exist almost everywhere in this region \citep{davi99}. The $^{12}$CO emission in SMM2A and SMM2B shows an elongation in the east-west direction passing SMM2A and a U-shaped structure surrounding SMM2B. These structures show a wide velocity range of $\sim 10\ \kms$. The morphology and velocity in the $^{12}$CO line may suggest effects of large-scale outflows in SMM2A and SMM2B. Such interaction among neighboring protostars should occur more easily in cluster environments than in isolated environments. The recent ODISEA 1.3 mm survey toward the moderately dense Ophiuchus star forming region also reported that only 23 of 133 Spitzer-selected protoplanetary disks have radii larger than 30 au \citep{ciez19}. This result along with our result in the clusters may suggest that cluster environments play a role to generate small disks.


\section{Conclusions} \label{sec:conc}
We have used ALMA to observe four submillimeter condensations, SMM2, SMM4, SMM9, and SMM11 in the star forming cluster Serpens Main, at an angular resolution of $\sim 0\farcs 55$ (240 au) in the 1.3 mm continuum, $^{12}$CO $J=2-1$ and C$^{18}$O $J=2-1$ lines. The main results are summarized below.
\begin{enumerate}
    \item We detected sixteen sources and divided them into three groups: six are associated with extended continuum emission and extended $^{12}$CO outflows (Group 1), four are associated with a filamentary structure in the 1.3 mm continuum and lack $^{12}$CO emission (Group 2), and six are unresolved (FWHM$\lesssim 120$ au) in the continuum and associated with clumpy $^{12}$CO outflows (Group 3).
    \item We interpret Group 1 as Class 0 protostars from the presence of outflows and their SEDs. The intensity-weighted lengths, orientation angles, opening angles, and velocities of the Group 1 outflows are measured from integrated intensity maps of the $^{12}$CO line. Their inclination angles are also estimated by fitting the wind-driven-shell model to the $^{12}$CO integrated intensity maps and position-velocity diagrams along the outflow axes, allowing us to estimate dynamical time of the outflows accurately. The dynamical time of the $^{12}$CO outflows, distribution of SO emission, C$^{18}$O fractional abundance, and continuum morphology suggest evolutionary trends within the Class 0 stage: SO jet in a young phase, CO desorption, and disk formation.
    \item The Group 2 members are not associated with protostars but are marginally Jeans unstable. Our interpretation of these are prestellar sources which is strengthened by an anticorrelation between density and C$^{18}$O emission.
    \item We interpret Group 3 as sources in the later stages of disk mass accretion based on their low mass envelopes and compact, faint $^{12}$CO outflows. Two Class 0 members of Group 3 show deconvolved radii of the disks $r<30$ au, whereas a Class 0 member of Group 1 shows a disk with $r\sim 240$ au.  The lack of a significant reservoir in the Group 3 members suggests that the disks are unlikely to substantially grow further around them. These results suggest that protostellar evolution depends on the initial, Class 0, conditions, and may explain the wide range of disk sizes in the T Tauri phase.
\end{enumerate}

\acknowledgments

We thank the anonymous referee, who gave us invaluable comments to improve the paper.
This paper makes use of the following ALMA data: ADS / JAO.ALMA2015.1.01478.S (P.I. Y. Aso). ALMA is a partnership of ESO (representing its member states), NSF (USA) and NINS (Japan), together with NRC (Canada), NSC and ASIAA (Taiwan), and KASI (Republic of Korea), in cooperation with the Republic of Chile. The Joint ALMA Observatory is operated by ESO, AUI/NRAO and NAOJ.
Y.A. acknowledges a grant from the Ministry of Science and Technology (MoST) of Taiwan (MOST 108-2112-M-001-048). N.H. acknowledges a grant from the Ministry of Science and Technology (MoST) of Taiwan (MoST 108-2112-M-001-017). Y.A. acknowledges a grant from NAOJ ALMA Scientific Research Grant Number of 2019-13B. S.T. acknowledges a grant from JSPS KAKENHI Grant Number JP18K03703 in support of this work. This work was supported by NAOJ ALMA Scientific Research grant No. 2017-04A. H.-W.Y. acknowledges support from MOST 108-2112-M-001-003-MY2. J.P.W. thanks the NSF for support through grant AST-1907486.

%

\vspace{5mm}
\facilities{ALMA}

\software{CASA \citep{mcmu07}, MIRIAD \citep{saul95}}

\end{document}